\documentclass[reprint, amsmath, amssymb, aps, prd, longbibliography, superscriptaddress]{revtex4-1}

\usepackage{listings}
\usepackage{booktabs}   %% For formal tables:
\usepackage{graphicx}
\usepackage{amsmath}
\usepackage{braket}
\usepackage{multirow}
\usepackage[colorlinks=true,urlcolor=blue,citecolor=blue,linkcolor=blue]{hyperref}
\usepackage{xcolor}
\usepackage[htt]{hyphenat}  %% Allow texttt to line break
\usepackage{booktabs}
\usepackage{yquant}
\useyquantlanguage{groups}
\usepackage{appendix} 

\usepackage[frozencache,cachedir=.]{minted}
\definecolor{mintedbackground}{rgb}{0.95,0.95,0.95}

\newmintedfile[qasmfile]{qasmlexer.py:QASMLexer -x}{
bgcolor=mintedbackground,
fontsize=\scriptsize,
linenos=true,
 }
 
 \newmintedfile[qasmfilewithlines]{qasmlexer.py:QASMLexer -x}{
bgcolor=mintedbackground,
fontsize=\scriptsize,
linenos=true,
}

\newminted[qasmblock]{qasmlexer.py:QASMLexer -x}{
bgcolor=mintedbackground,
fontsize=\scriptsize,
linenos=true,
} 

\newminted[pythonblock]{python}{
bgcolor=mintedbackground,
fontsize=\scriptsize
} 

\providecommand{\ket}[1]{\ensuremath{\left|#1 \right\rangle}}

\begin{document}

\title{Quantum computing with Qiskit}
\author{Ali Javadi-Abhari}
\affiliation{IBM Quantum, IBM T. J. Watson Research Center, Yorktown Heights, NY, 10598}

\author{Matthew Treinish}
\affiliation{IBM Quantum, IBM T. J. Watson Research Center, Yorktown Heights, NY, 10598}

\author{Kevin Krsulich}
\affiliation{IBM Quantum, IBM T. J. Watson Research Center, Yorktown Heights, NY, 10598}

\author{Christopher J. Wood}
\affiliation{IBM Quantum, IBM T. J. Watson Research Center, Yorktown Heights, NY, 10598}

\author{Jake Lishman}
\affiliation{IBM Quantum, IBM Research Europe, Hursley, United Kingdom}

\author{Julien Gacon}
\affiliation{IBM Quantum, IBM Research Europe, Z{\"u}rich, Switzerland}

\author{Simon Martiel}
\affiliation{IBM Quantum, IBM France Lab, Orsay, France}

\author{Paul D. Nation}
\affiliation{IBM Quantum, IBM T. J. Watson Research Center, Yorktown Heights, NY, 10598}

\author{Lev S. Bishop}
\affiliation{IBM Quantum, IBM T. J. Watson Research Center, Yorktown Heights, NY, 10598}

\author{Andrew W. Cross}
\affiliation{IBM Quantum, IBM T. J. Watson Research Center, Yorktown Heights, NY, 10598}

\author{Blake R. Johnson}
\affiliation{IBM Quantum, IBM T. J. Watson Research Center, Yorktown Heights, NY, 10598}

\author{Jay M. Gambetta}
\affiliation{IBM Quantum, IBM T. J. Watson Research Center, Yorktown Heights, NY, 10598}

\newcommand{\ali}[1]{\textcolor{red}{AJ: #1}}
\newcommand{\jay}[1]{\textcolor{blue}{JG: #1}}
\newcommand{\matthew}[1]{\textcolor{green}{MT: #1}}

\begin{abstract}

We describe Qiskit, a software development kit for quantum information science. We discuss the key design decisions that have shaped its development, and examine the software architecture and its core components. We demonstrate an end-to-end workflow for solving a problem in condensed matter physics on a quantum computer that serves to highlight some of Qiskit's capabilities, for example the representation and optimization of circuits at various abstraction levels, its scalability and retargetability to new gates, and the use of quantum-classical computations via dynamic circuits. Lastly, we discuss some of the ecosystem of tools and plugins that extend Qiskit for various tasks, and the future ahead.

\end{abstract}

\maketitle

\twocolumngrid

\section{Introduction}\label{sec:intro}

Quantum computing is progressing at a rapid pace, and robust software tools such as Qiskit are becoming increasingly important as a means of facilitating research, education, and to run computationally interesting problems on quantum computers. For example, Qiskit was used in a recent paper that showed evidence of utility for quantum computers by using error mitigation~\cite{kim2023evidence}. It was also used in a demonstration of fault-tolerant magic state preparation beyond break-even fidelity~\cite{gupta2023encoding}, and in a number of utility-scale papers comprised of up to 133-qubits and $\mathcal{O}(10^{4})$ two-qubit entangling gates \cite{zhang:2023,majumdar:2023,yu2023simulating,shtanko:2023,baumer:2023,chen:2023,yasuda:2023,farrell2023scalable,pelofske:2023,farrell:2024,shinjo:2024,miessen:2024,robledo:2024,montanez:2024}.

Qiskit was started in 2017 as an open-source toolbox for quantum computing by IBM. Six years after its initial release, the Qiskit ecosystem is thriving. The package has been installed over 6 million times, at a current rate of 300,000 per month. 500\textsuperscript{+} individuals have contributed to its development, the vast majority of whom are unaffiliated with IBM. 300 packages in the Python Package Index (PyPI) depend on Qiskit, as well as hundreds of other code repositories on Github~\cite{github,pypi}. More than 2,000 scientific papers posted to arXiv have used Qiskit, and many university courses have used Qiskit in their material. By a large margin, Qiskit is the most widely-adopted quantum computing software~\cite{unitaryfund}.

Recently Qiskit reached a 1.0 major release milestone. In this paper we present Qiskit's overall design philosophy and software architecture, with a deeper dive into circuits, pass managers and primitives, which form its core components. We showcase some of its capabilities through an end-to-end workflow where a Hamiltonian simulation problem is solved using Qiskit on a quantum computer. We demonstrate how Qiskit can be leveraged to improve the experimental performance using a variety of techniques, including by targeting hardware-native gatesets, leveraging dynamic circuits and suppression of errors. Finally we discuss the software ecosystem that has formed around Qiskit, and the future ahead.

%The rest of this paper is organized as follows: Section~\ref{sec:design} discusses the main design decisions and the driving philosophy behind Qiskit. Section~\ref{sec:arch} gives an overview of the software architecture, with a deeper dive into circuits, operators, pass managers and primitives, which form the core components of Qiskit. We combine these in Section~\ref{sec:examples} to show an example workflow where a Hamiltonian simulation problem is solved using Qiskit on a quantum computer. We demonstrate how Qiskit can be leveraged to improve the experimental performance using a variety of techniques, including by targeting hardware-native gatesets and leveraging dynamic circuits. Section~\ref{sec:ecosystem} illustrates some software tools from the community that have built on Qiskit, and Section~\ref{sec:conclusion} concludes the paper with a look towards the future.
\section{Design Philosophy}\label{sec:design}

\begin{figure*}
\includegraphics[width=\textwidth]{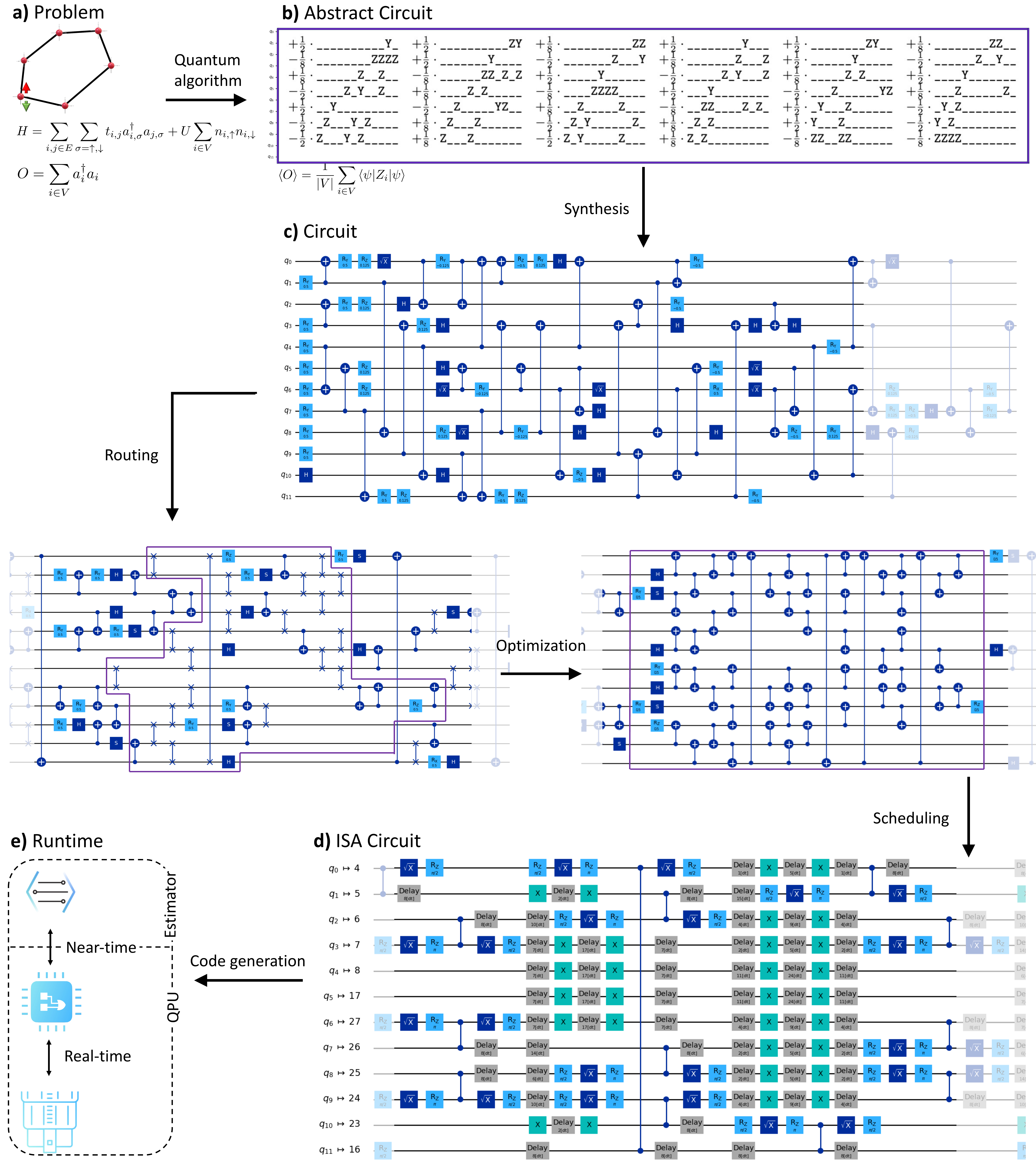}
  \caption{The quantum computing software stack at various abstraction levels. Qiskit's scope spans the middle-end from (b) to (d), which includes all the tools and algorithms needed to efficiently represent and compile circuits.
  a) A representation of the problem and the observable to be measured, for example the dynamics of a Fermi-Hubbard model. b) Translation of the problem to an abstract Hamiltonian simulation circuit, represented by exponentials of Pauli operators and an observable represented by Pauli operators. c) Synthesis of the high-level operators to a circuit with gates from a standard library. The circuit is then transpiled through a set of rewriting stages: routing, optimization and scheduling each transform the circuit to comply with the target ISA and suppress errors. Here an example peephole optimization is illustrated that optimizes a depth-26 subcircuit (counted by CNOTs) to a depth-12 subcircuit. d) ISA circuit that is compatible with the hardware's instruction set, where all gates and their schedule is explicit. The ISA is a feature of the underlying quantum computer, and differs among various technological platforms and between fault-tolerant vs. pre-fault-tolerant quantum computers. e) Classical code generation for a runtime environment and classical controller to execute the ISA circuit on the quantum device.
  }
  \label{fig:stack}
\end{figure*}

We begin by discussing Qiskit's scope within the broader quantum computing software stack, as illustrated in Figure~\ref{fig:stack}. Starting from a computational problem, a quantum algorithm specifies how the problem may be solved with quantum circuits. This step involves translating the classical problem to the quantum domain, for example Fermion to qubit mapping~\cite{qiskit-nature, mcclean2020openfermion}. Circuits at this level can be quite abstract, for example only specifying a set of Pauli rotations, some unitaries, or other high-level mathematical operators. Importantly, these abstract circuits are representable in Qiskit, which contains synthesis methods to generate concrete circuits from them. Such concrete circuits are formed using a standard library of gates, representable using intermediate quantum languages such as OpenQASM~\cite{cross2022openqasm}. 

The transpiler rewrites circuits in multiple rounds of passes, in order to optimize and translate it to the target instruction set architecture (ISA). The word ``transpiler'' is used within Qiskit to emphasize its nature as a circuit-to-circuit rewriting tool, distinct from a full compilation down to controller binaries which is necessary for executing circuits. But the transpiler can also be thought of as an optimizing compiler for quantum programs.

The ISA is the key abstraction layer separating the hardware from the software, and depends heavily on the quantum computer architecture beneath. For example for a physical quantum computer based on superconducting qubits, this can include $\tt CNOT$, $\tt \sqrt{X}$ and $\tt RZ(\theta)$ rotations. For a logical quantum computer, it can include joint Pauli measurements, magic state distillation, or other operations specific to the error correcting code~\cite{beverland2022assessing}. Note that the ISA is often more than just a universal set of quantum gates, and can include {\tt measure}, {\tt reset} or {\tt delay} operations, or classical control-flow such as {\tt if/else} branches and {\tt for/while} loops, or limited concurrent classical computations on classical data. For example see Appendix~\ref{sec:isa} for the current ISA pertaining to IBM's quantum computers.

In the rest of this section we discuss some key design decisions that have informed Qiskit's development.

\subsubsection{Modularity and extensibility}
Quantum computing is an active area of research~\cite{huang2020predicting}, and major advances in many areas are needed to solve interesting and classically intractable problems. 
Qiskit is built as a tool to explore quantum computing and drive innovation. As such, modularity and extensibility are principal driving forces behind the software design. For example, it is straightforward to extend its library of circuits or circuit synthesis and optimization methods. These can be completely outside of Qiskit, visible to Qiskit only via a plug-in interface, so that research code can be seamlessly used by Qiskit users. In Section~\ref{sec:ecosystem} we give examples of some projects that extend Qiskit in interesting ways using this mechanism.

\subsubsection{Balance between performance and rapid prototyping}
Speed is a fundamental consideration for large-scale quantum computing, and software performance must not hinder the overall workflow. As quantum computers scale, larger and larger circuit volumes must be analyzed and transformed rapidly. In addition, many quantum algorithms as well as error mitigation protocols require a large number of circuit executions~\cite{peruzzo2014variational, van2023probabilistic}. As such, Qiskit places a significant focus on speed. Performance-critical components and algorithms are written in the Rust programming language~\cite{10.1145/2663171.2663188}, and performance is tracked over time using extensive benchmarks~\cite{asv}.  However, Python remains the language of choice for many in the scientific community due to its ease of adoption and lower barrier of entry. By moving only the performance-critical aspects of the code to Rust (currently about 6\% of the code base), Qiskit maintains a Python environment for programming and prototyping, striking a balance between speed and ease of use.

\subsubsection{Balance between portability and hardware optimization}
Qiskit uses universal circuit representations and transformations that are agnostic to the underlying hardware. This enables writing quantum programs in a natural manner, without having to worry about implementation details. However, Qiskit is also able to transform circuits to make them compatible with diverse quantum platforms with different instruction set architectures, such as superconducting or trapped-ion technologies~\cite{qiskit-ibm-runtime, qiskit-ionq, qiskit-rigetti}.
Qiskit can represent many types of quantum hardware through its {\tt Target} class, which is an abstract machine model. The {\tt Target} defines a model for describing the instructions available on quantum hardware, their properties such as timings and error rates, and other constraints of the hardware. Qiskit's transpiler is retargetable, and can use this information to optimize the circuit for a given hardware.

\subsubsection{Interoperability across abstraction levels}
Another goal in Qiskit's evolution has been to give more control to users by allowing them to program at different abstraction levels. A high level programming model is attractive since it allows users to focus on code development and not worry about the details that go into realizing a given computation on a physical machine. Other users may want to investigate the physics behind quantum computing, through timing and dynamics of gates. All of these abstraction levels are compatible with each other in the same Qiskit circuit, and optimizations that cut across these layers are often employed to save on resources, for example direct synthesis of high-level operators for a given Target, without intermediary steps that may prolong circuit depth.

\subsubsection{Quantum-classical integration: real time and near time}
Quantum computation is often more than just a unitary time evolution of quantum states. Classical resources must be tightly integrated with quantum computers to enable different kinds of computation, such as using classical control flow to extend the computational reach of quantum circuits~\cite{bravyi2020quantum} or to correct errors~\cite{bravyi2005universal}, or using classical computation to optimize the parameters of a quantum circuit~\cite{peruzzo2014variational}

We distinguish between two types of classical computation. The first are ``real-time'' classical computations that occur while qubits are coherent, such as control flow based on the outcome of qubit measurements. We call circuits of this type {\em dynamic circuits}, as the instructions that are executed are determined dynamically as the circuit progresses. Qiskit's circuit model allows for a rich mixture of classical operations concurrent with quantum operations, similar to the circuit model defined in the OpenQASM 3~\cite{cross2022openqasm} language.

The second type of classical computation occurs in a ``near-time'' environment, which does not have the same stringent timing requirements, but must still occur with low latency. Examples include parameter binding, just-in-time compilation, optimization of circuit parameters, or error mitigation. Qiskit is designed to be a lightweight framework that can be integrated into a runtime environment that co-locates quantum with general-purpose classical processors, so that thousands of circuits can be rapidly generated and evaluated in a dynamic manner to provide the final solution. An example of this runtime environment is the Qiskit Runtime implemented in IBM quantum computers.

\subsubsection{Computational primitives} 
Beside specifying the quantum circuit, the computation's output is also a key consideration. In quantum computing there exist two main primitives for capturing the output of a quantum circuit: sampling output bitstrings, or estimating observable expectation values. These primitives are the means by which circuits are evaluated in Qiskit. They define a common interface, even if the implementation details can vary significantly. For example, efficient estimators are an active area of research~\cite{huang2020predicting, elben2023randomized}, as are error mitigation methods that can improve their results~\cite{van2023probabilistic, kim2023evidence}. Nonetheless, they all follow a simple interface of taking a (possibly parameterized) ISA circuit, coupled with one or more observables, and returning estimates of their expectation values. Different estimators can thus be exposed to Qiskit as a primitive, i.e. effective executors for quantum computers.

\subsubsection{Qiskit patterns}
A common flow for using Qiskit is through a four-step workflow, and the software architecture reflects this (see Figure~\ref{fig:architecture}). First, a classical problem is mapped to quantum computation by generating circuits that encode the problem. This step is best handled by domain-specific software or experts, although Qiskit provides a convenient circuit construction API that can handle large circuits. Next, the circuits are transformed to make them amenable for execution on a target hardware. We generically refer to this step as transpilation, as it is a circuit-to-circuit rewriting step, and not a full compilation down to the classical controller instructions. Next, the circuits are evaluated using primitive computations on a target backend. Finally, the results are post-processed to obtain a solution to the original problem. 

Actual workflows may iterate through a number of such steps, and we envision multiple patterns of this type which we call a Qiskit pattern. For example, a common pattern is to iteratively generate new circuits that depend on the results obtained from a previous batch of circuits, which can be seen as a loop around the last three steps above~\cite{cerezo2021variational}. More advanced patterns may leverage quantum compute within large-scale classical compute in a quantum-centric supercomputing architecture~\cite{alexeev2024quantum, robledo:2024}. The orchestration of complex patterns is simplified through the Qiskit serverless framework~\cite{faro2023middleware}.

\subsubsection{Integrated visualizations}
A popular feature of Qiskit are its visualization capabilities. This includes advanced circuit visualizers that display large circuits of varying characteristics (e.g. see Figure~\ref{fig:circuits}), target visualizer for giving an overview of the hardware topology and supported gates, and state and distribution visualizers for understanding computational outputs. Other types of information useful for debugging, such as the sequence of passes in the transpiler or the graph structure of the circuit can also be visualized.

\subsubsection{Tensor ordering convention}
When interpreting circuits, a convention must be picked for the ordering of qubits in a register, arguments of all instructions, and of instructions themselves. For example it is common that instructions in a circuit are ordered from left to right (i.e. temporal ordering), which is the opposite of how gate matrices are multiplied in the circuit.

Throughout Qiskit, tensor products are ordered as $Q_2 \otimes Q_1 \otimes Q_0$. For example the Pauli $ZX$ means $Z_1X_0$.
In this convention, the computational basis can be conveniently represented as $\ket{000}=\ket{0}$, $\ket{001}=\ket{1}$, $\ket{010}=\ket{2}$, $\ket{011}=\ket{3}$, etc. This leads to more natural representations of reversible circuits too, as numbers are encoded in qubit registers in the same way as they would be in classical registers.  Note that this is merely a convention, which may be natural for some tasks but not others. For example, the qubit arguments to a $CX$ instruction are ordered as $Q_0=\text{control}$  and $Q_1=\text{target}$, This leads to a unitary of $CX = \begin{bmatrix}
1 & 0 & 0 & 0 \\
0 & 0 & 0 & 1 \\
0 & 0 & 1 & 0 \\
0 & 1 & 0 & 0
\end{bmatrix}$, which may be different to some textbook definitions of the gate~\cite{nielsen2001quantum}. The unitary can be changed, however, by simply using it in the circuit context as $CX (Q_1, Q_0)$. Similar qubit ordering conventions have been adopted elsewhere in the literature~\cite{smith2017someone, quirk}.

\section{Software Architecture}\label{sec:arch}

\begin{figure*}
  \includegraphics[width=\textwidth]{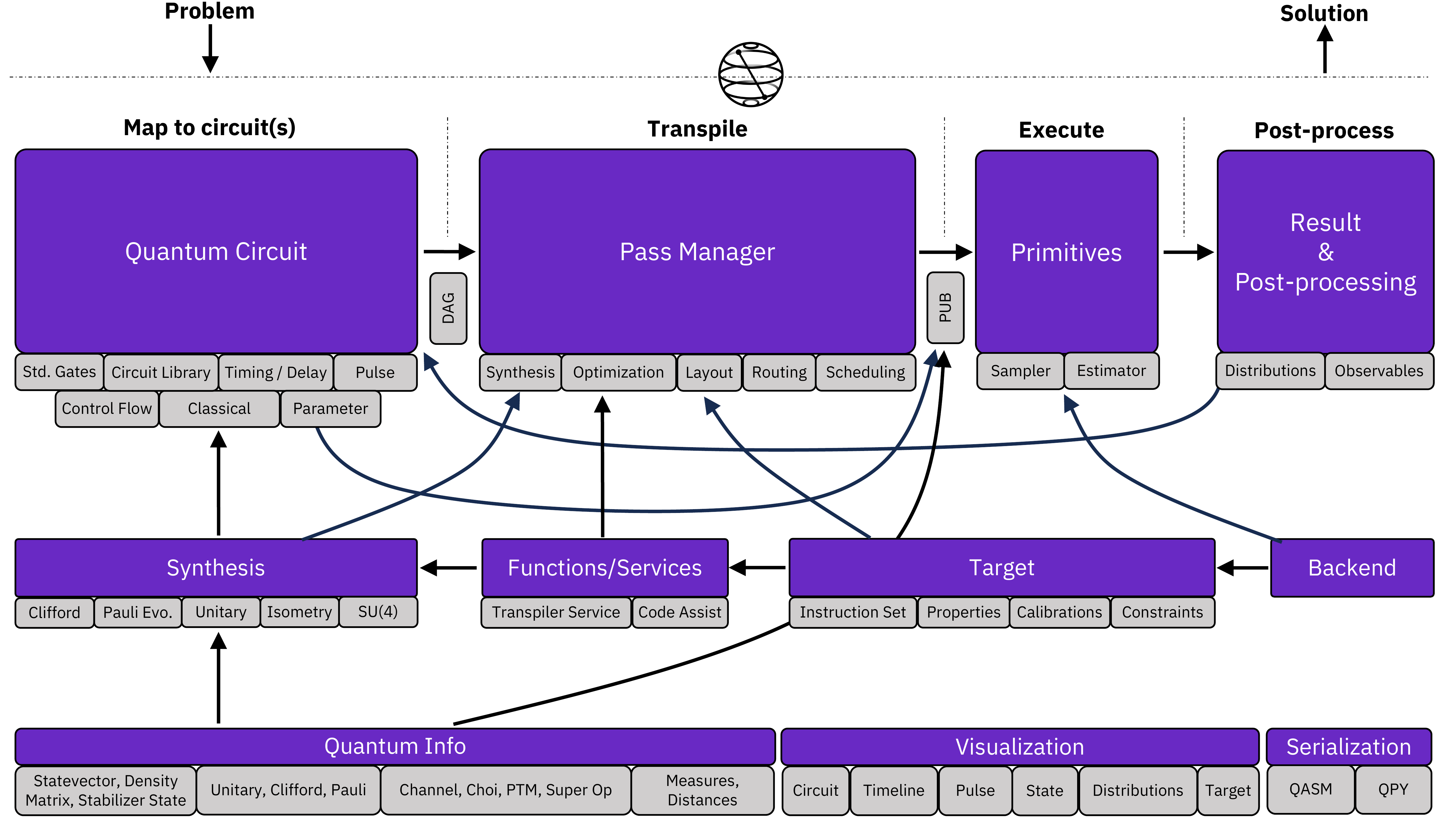}
  \caption{Qiskit's software architecture. The core component in Qiskit is the quantum circuit, around which the rest of the framework revolves. The quantum info module serves as a toolbox to connect circuits to the mathematical formalism of quantum information. Beside tools for representing channels and measuring fidelities, it includes operators that can be used in building abstract circuits, or be used for observable measurements. The transpiler transforms quantum circuits by applying a pipeline of passes on them, orchestrated by a pass manager. Circuit transformations occur according to a target, which is an abstract machine model that summarizes the pertinent features of a backend for the pass manager, such as its instruction set architecture (ISA) and any properties or constraints associated with it. Primitives evaluate circuits via backends, which may consist of simulators or hardware. Finally, circuits are evaluated with respect to some primitives, such as observable expectation values or measurement samples. These can be post-processed to obtain a solution, or be used to generate a next set of circuits. Visualization capabilities are supported throughout. Serialization via the OpenQASM language~\cite{cross2022openqasm} is used for execution of ISA circuits, whereas QPY-format serialization is used with Qiskit-native objects for use with functions and services~\cite{kremer2024practical}.
}
  \label{fig:architecture}
\end{figure*}

Figure~\ref{fig:architecture} shows an overview of Qiskit's architecture, including key components and how they interact. The backbone of the software consists of circuits, pass managers and primitives, which we describe next.

\subsection{Circuits}
Quantum circuits form the core component of Qiskit. Circuits express a particular computation as a time-ordered set of instructions, which can then be transformed and analyzed by the rest of the software.

Circuits in Qiskit are defined very broadly --- any operation on quantum or classical data can be included in the circuit. This includes standard operations such as qubit resets, gates and measurements, but also higher-level mathematical operators such as unitaries, isometries, Cliffords or Fourier transforms. Circuits may contain real-time classical computation on classical data while qubits are coherent, such as Boolean functions applied to the outcome of measurements, as well as real-time classical control flow like loops and branching. Qiskit circuits may also describe the timing of operations and even continuous-time dynamics of qubits via pulse-defined gates. Any of these levels of abstraction may be mixed and matched within the same circuit, and circuits can be composed with each other like building blocks. We use these different types of circuits in Section~\ref{sec:examples} to solve the same problem in a variety of ways.

\begin{figure*}
  \includegraphics[width=\textwidth]{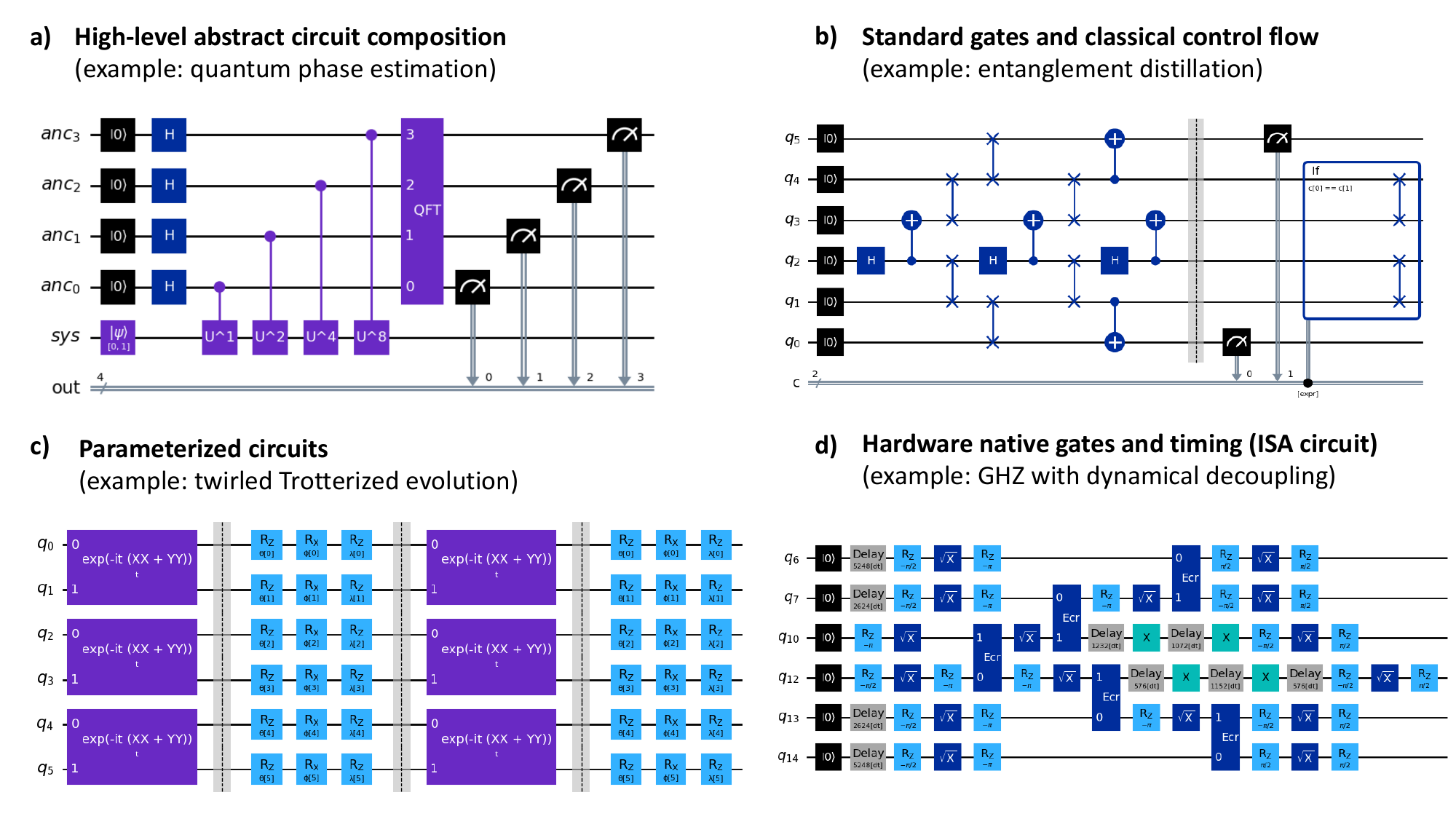}
  \caption{Examples of different kinds of circuits in Qiskit. All circuits are visualized using Qiskit.
  a) A quantum phase estimation algorithm~\cite{kitaev1995quantum}. The circuit is built using standard gates as well as some high-level blocks: state preparation $\ket{\psi}$, unitaries raised to a power and controlled, and quantum Fourier transform. The transpiler decides how to synthesize the circuit efficiently from this high-level representation.
  b)An entanglement distillation circuit that performs a pipelined preparation of three Bell states using gates from the standard library. Two of the pairs are distilled into one~\cite{bennett1996mixed}, and in the event that the distillation fails, the protocol falls back on the third pair, thereby avoiding post-selection failure. This latter operation is described by a classical {\tt if} statement on measured qubit outcomes. 
  c) A Trotterized $XX+YY$ Hamiltonian simulation~\cite{lloyd1996universal} circuit, with each layer's coherent noise suppressed by Pauli twirling~\cite{bennett1996purification}. This light-weight circuit pattern can turn into many circuit instances on the fly by substituting different time evolution parameters or angle parameters to sample different twirls. 
  d) A GHZ state preparation circuit. This circuit is lowered to the native instruction set architecture (ISA) of a hardware where instruction durations are known from calibrations. The precise timing allows control techniques such as dynamical decoupling~\cite{viola1999dynamical} to be applied that refocus the noise in hardware.
  } 
  \label{fig:circuits}
\end{figure*}

Figure~\ref{fig:circuits} shows some examples of the various types of circuits that can be described in Qiskit. This flexibility enables the study of a wide variety of quantum algorithms and physical implementations. Owing to the flexibility of the circuit data structure, it is easy to extend it beyond its default scope, and we illustrate some examples of this in Section~\ref{sec:ecosystem}.

Distinct from how users interact with circuits in Qiskit, the internal representation of circuits can take multiple forms, each suitable for specific purposes. The default data structure is a list of instructions (i.e.\ operations applied to quantum or classical data). However, in many circuit rewriting algorithms a data-flow graph structure, represented by a directed acyclic graph (DAG) is more suitable, since it makes the succession of operations and flow of information explicit. Likewise, some circuit rewriting algorithms benefit from a canonical graph form where gate commutations are taken into account and only true dependencies between instructions are encoded. Particularly structured sub-circuits may also be temporarily converted to specialized data structures to facilitate reasoning about the circuit at a more abstract level, for example Boolean linear functions~\cite{patel2008optimal}, Cliffords~\cite{aaronson2004improved}, or phase polynomials~\cite{amy2014polynomial}
.

Given the quantum circuit's central role, a particular emphasis in these data structures is to make circuits as light weight as possible. For example, certain common gates are defined as singletons in the standard library, blocks can be reused in circuits, and circuit synthesis and lowering for abstract operations is lazily deferred to the transpiler. This ensures that the software remains scalable as increasingly larger quantum computations are studied.

\subsection{Pass Manager}
The Qiskit transpiler contains a collection of passes implementing proven translation and optimization techniques on top of a flexible framework for describing, composing and running pipelines of quantum circuit transformations. This enables authors of both high-level abstract quantum circuits and low-level hardware-aware quantum circuits to benefit from both Qiskit's included passes to generate a device-compatible and device-optimized implementation of their input circuit as well as an API for automating circuit transformations.

As Qiskit circuits support multiple levels of abstraction, although the circuit is progressively lowered and transformed, the output representation of the transpiler is also a quantum circuit.
This architecture promotes inspection, characterization and modification of transpiled circuits using the same tooling available for circuit construction, aiding users in understanding their circuit's execution cost relative to a finite error budget, and facilitates experimentation with alternative optimization techniques which may bring advantages for specific applications or device architectures to take maximal advantage of modern device capabilities.

Underlying the transpiler is a pass-based infrastructure called the Pass Manager which supports the construction and manages the execution of composable and re-usable pipelines of semantic-preserving transformations for quantum circuits, including logic for controlling the flow of the compilation pipeline depending on device or circuit characteristics, or properties of the intermediate compilation steps. This architecture is similar to classical compiler infrastructures such as LLVM~\cite{lattner2004llvm}.

Qiskit defines a series of standardized stages of compilation, each comprised of one or more transformation and analysis passes, which in turn define standardized hook points for transpiler plugins and extensions to modify or extend the predefined Pass Managers defined and distributed with Qiskit.
These Pass Managers encapsulate best-practices in circuit optimization, translation and re-writing for a broad set of applications and hardware architectures, and serve as the basis for full-pipeline compiler performance comparisons for experimentation and extensions.

\subsection{Primitives}

Finally, circuit executions are performed using primitives, which provide a consistent API for performing common quantum computational tasks. In particular, the primitives consist of a sampler and estimator. Samplers return measurement outcomes, while estimators return post-processed expectation values. 

The primitives can be implemented in a Qiskit runtime environment, which provides a tightly-coupled interface of quantum-classical near-time compute. Note that this is distinct from real-time classical compute which must occur at a much higher speed and is consequently much more limited in scope. By co-locating general-purpose classical computers with the quantum processor, a runtime environment can cut down the latency on the execution of quantum-classical workloads. This could consist of optimizing circuits to suppress errors, generating and running extra circuits for error mitigation, post-processing results or updating parameters.

% The runtime architecture provides multiple benefits. 
% First, by co-locating classical with quantum compute, it cuts down on the latency of cloud communication. In many scenarios, such as the calculation of error-mitigated expectation values, many intermediary execution results will be generated which need not be communicated to the end user. Furthermore, the quantum processor may be used interactively, for example to improve precision. All of these can be done within the runtime.
% Second, the Qiskit Runtime enables the holding of state across multiple invocations, or even across different sessions. Examples include circuit transpilations that are only done once and thereafter updated in parameters, or learned noise models that can persist over time.
% Third, user code can be containerized in the runtime for security, yet trusted programs such as primitives can have privileged access to the quantum processor for speed. This access allows uninhibited use of the QPU and shared memory to improve execution speed.
% Fourth, the co-located hardware can be extended as needed to accommodate special needs, such as applying optimal control or error correction.

\section{Qiskit by Example}\label{sec:examples}

We now show an end-to-end example that uses Qiskit to solve a problem in condensed matter physics on a quantum computer. The purpose is to highlight general workflows for solving problems using Qiskit, and some of the software's capabilities. Specifically, we consider the kicked Ising model for a lattice of spins. This problem is widely studied in statistical mechanics and was the subject of the large scale ``quantum computer utility'' experiment performed in~\cite{kim2023evidence}. In one instance, we show how changing the target to include new calibrated gates automatically prompts the transpiler to re-target and re-optimize the circuit to reduce depth. In another instance, we show how Qiskit's dynamic circuit capabilities can make shorter and more reliable circuits especially when faced with limited connectivity. %Throughout, we show concrete steps for defining circuits, observables and parameters of a given problem, transpiling circuits, and using primitives to measure observables on a quantum computer.

\subsection{Ising model simulation on a quantum computer}

\begin{figure*}
  \includegraphics[width=\textwidth]{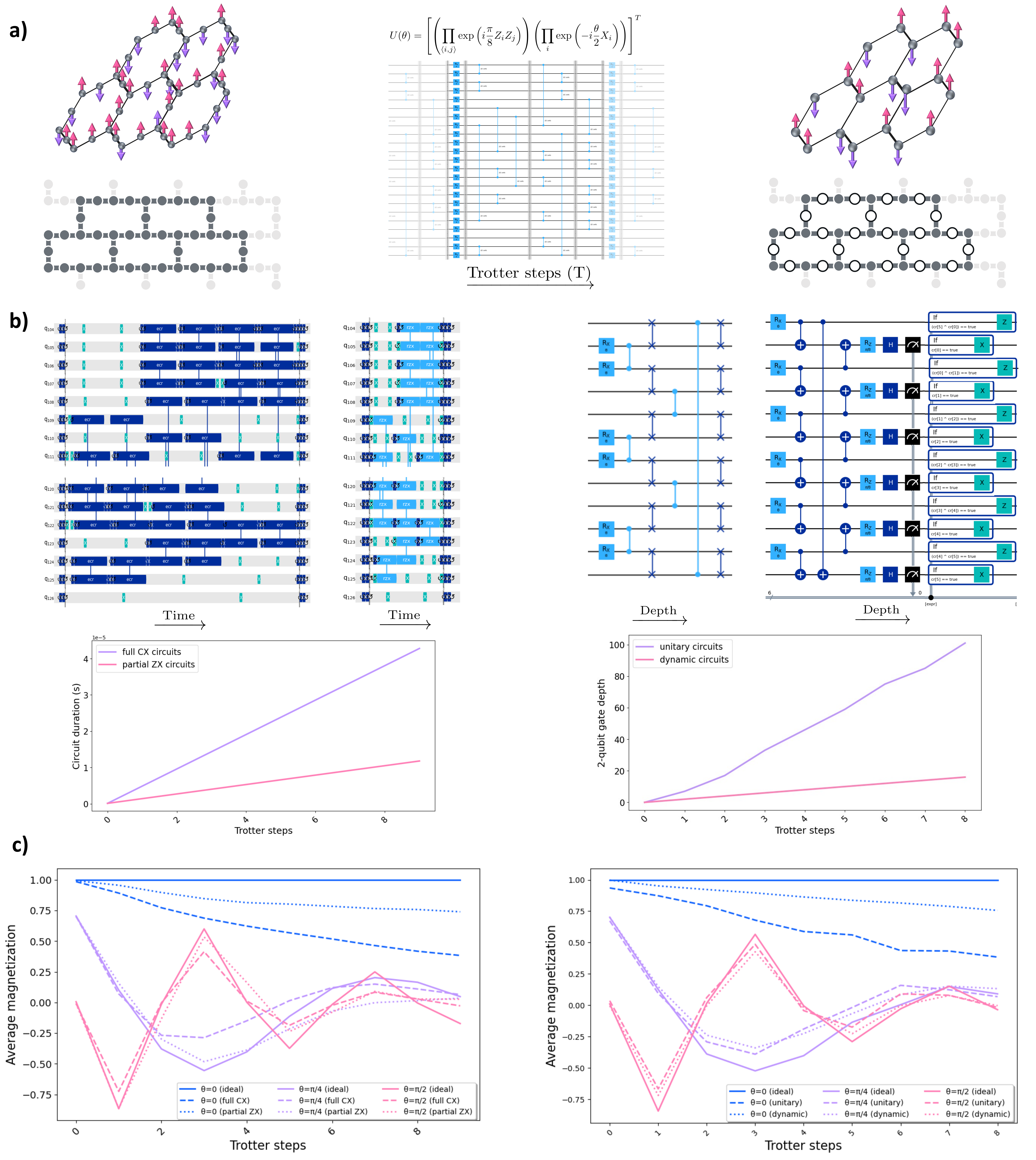}
  \caption{Using Qiskit to solve a Hamiltonian simulation problem on a quantum computer. a) Building a (parameterized) circuit to represent the problem. We consider a heavy-hexagonal as well as honeycomb lattice of spins. The former matches the lattice of hardware qubits, while the latter does not. The Hamiltonian simulation consists of multiple Trotter steps that alternate $X(\theta)$ rotations with $ZZ(\pi/8)$ interactions on the edges of the lattice. Throughout this figure, the left (right) side corresponds to the heavy-hexagonal (honeycomb) lattice. Experimental results are for a 23-qubit (12-qubit) instance of the heavy-hexagonal (honeycomb) problem.
  b) Four circuits post-transpilation, each realizing a single Trotter step. Left: the transpiler lowers the circuit to the hardware ISA and inserts dynamical decoupling. The availability of new, shorter gates will prompt Qiskit to retarget the circuit to shorten its duration. Right: the transpiler uses SWAP gates to map the honeycomb lattice to the hardware. When we use dynamic circuits, we gain a reduction in depth.
  c) Experimental results using Qiskit primitives. Here we use the Estimator to obtain expectation values corresponding to the average magnetization of the spin system. The Estimator is invoked with built-in readout error mitigation, and $\theta$ parameters are substituted at runtime. Left: we see an improvement by targeting partial ZX gates compared to full CX gates. Right: we see an improvement by using dynamic circuits as opposed to routing with SWAPs.
  } 
  \label{fig:ising}
\end{figure*}

Let us begin by describing the overall problem and the setting for our experiment, which is also depicted in Figure~\ref{fig:ising}(a). The Hamiltonian that we simulate is as follows, where the system undergoes Ising interactions between lattice sites, as well as a local magnetic field on each site.

\begin{equation*}
    U(\theta) = \left( \prod_{\langle j, k \rangle} \exp\left( i\frac{\pi}{8} Z_{j} Z_{k} \right) \right) \left( \prod_{j} \exp\left( -i\frac{\theta}{2} X_{j} \right) \right)
\end{equation*}

The quantum computer that we use in this experiment is \emph{ibm\_pinguino1}, which has 127 fixed-frequency transmon qubits arranged in a heavy-hexagonal architecture, with microwave-controlled cross-resonance interactions between neighboring qubits. 

\subsection{Scalablility}

In the first experiment, shown in Figure~\ref{fig:ising}(b), we simulate a lattice which is identical to our heavy-hex qubit lattice, which allows us to study the problem at scale. Here we rely on Qiskit's scalablity to the regime of large circuits, and many circuits: the representation and processing of circuits containing many qubits and many gates is optimized through Qiskit's use of memory-efficient circuit representation and internal use of Rust for transpilation. In addition, when the goal is to evaluate many similar circuits using a sweep of parameters, Qiskit's parameter binding framework allows lower-level hardware to efficiently process the same circuit template with many different parameter substitutions. Multiple parameter bindings to a common circuit template is a common task in quantum computing, and is therefore explicitly provisioned for in the primitives interface.

We begin by obtaining the IBM Quantum backend on which we define our lattice problem.

\begin{pythonblock}
from qiskit_ibm_runtime import QiskitRuntimeService

service = QiskitRuntimeService()
backend = service.backend('ibm_pinguino1')
\end{pythonblock}

We pick a 23-qubit subgraph of the hardware qubits, and use the RustworkX graph library~\cite{treinish:2021} to edge-color the subgraph. Each color corresponds to one (non-intersecting) set of edges whose qubit pairs can simultaneously interact.

\begin{pythonblock}
import rustworkx as rx

target = backend.target
coupling_map = target.build_coupling_map()
G = coupling_map.graph.to_undirected(multigraph=False)
qubit_subset = list(range(104, 127))
subgraph = G.subgraph(qubit_subset)
qubit_map = dict(enumerate(sorted(qubit_subset)))
edge_colors = rx.graph_misra_gries_edge_color(subgraph)
layer_edges = {color: [] for color in edge_colors.values()}
for edge_index, color in edge_colors.items():
    src, tgt = subgraph.edge_list()[edge_index]
    layer_edges[color].append((qubit_map[src], qubit_map[tgt]))
\end{pythonblock}

We now define our Trotterized time-evolution circuit. At each Trotter step there are three layers of two-qubit $ZZ$ interactions corresponding to the previously-defined edge colors, as well as one layer of single-qubit $X$ rotations. We consider between 0 to 9 Trotter steps. We choose a $\pi/8$ angle for our $ZZ$ interactions, and leave the $X$ rotations parameterized. We later bind these angles at runtime to give us a picture of the Ising time evolution under different kick strengths. 

\begin{pythonblock}
# step 1: build circuit for problem

import numpy as np
from qiskit.circuit import QuantumCircuit, Parameter

num_qubits = backend.num_qubits
steps = range(10)
zz_angle = np.pi/8
theta = Parameter('theta')

circuits = []
for step in steps:
    circuit =  QuantumCircuit(num_qubits)
    for _ in range(step):
        for i in qubit_subset:
            circuit.rx(theta, i)
        circuit.barrier(*qubit_subset)
        for same_color_edges in layer_edges.values():
            for e in same_color_edges:
                circuit.rzz(zz_angle, e[0], e[1])
        circuit.barrier(*qubit_subset)
    for i in qubit_subset:
        circuit.rx(theta, i)
    circuits.append(circuit)
\end{pythonblock}

Next we define the observables that we wish to estimate from the circuits, which here we define as the average magnetization of all spins $\sum_{i=1}^{N}{\frac{1}{N}{Z_i}}$. We also specify the circuit parameters ($X$-rotation angles) that we wish to sweep.

\begin{pythonblock}
# step 1 (cont.): build observables and parameters

from qiskit.quantum_info import SparsePauliOp
obs = SparsePauliOp.from_sparse_list(
    [("Z", [i], 1/len(qubit_subset)) for i in qubit_subset],
    num_qubits=num_qubits)

max_angle = np.pi/2
points = 3
params = np.linspace(0, max_angle, points) 
\end{pythonblock}

Having defined our circuits and target hardware, we now use Qiskit's pass manager to transform the circuit to be executable on the hardware. Additionally, we modify the ``scheduling'' stage of the pass manager to insert dynamical decoupling sequences that ensure some noise suppression during execution.

\begin{pythonblock}
# step 2: transpile

from qiskit.transpiler import PassManager
from qiskit.transpiler.preset_passmanagers import (
    generate_preset_pass_manager
)
from qiskit.transpiler.passes import (ALAPScheduleAnalysis,
                                      PadDynamicalDecoupling)
from qiskit.circuit.library import XGate

pm = generate_preset_pass_manager(
    target=target, optimization_level=3)
pm.scheduling += PassManager([
    ALAPScheduleAnalysis(target=target),
    PadDynamicalDecoupling(dd_sequence=[XGate(), XGate()],
                           target=target)
])
circuits_isa = pm.run(circuits)
\end{pythonblock}

Finally, we can use the estimator primitive to estimate observable expectation values from our circuits, using the backend as the computational resource. Each unit of work done by the estimator contains one circuit and possibly multiple observables and parameters, which is also called a primitive unified block (PUB).

\begin{pythonblock}
# step 3: execute using primitives

from qiskit_ibm_runtime import EstimatorV2, EstimatorOptions

options = EstimatorOptions(optimization_level=0,
                           resilience_level=1)
estimator = EstimatorV2(backend=backend, options=options)
job_exp = estimator.run([(circ, [obs], params)
                         for circ in circuits_isa])
expvals_exp = [job_exp.result()[i].data.evs
               for i in range(len(steps))]
\end{pythonblock}

This gives the average magnetization in the Ising model for different number of Trotter steps and different kick strengths, which we can plot.

\begin{pythonblock}
# step 4: analyze results

import matplotlib.pyplot as plot
import matplotlib as mpl

plt.figure(figsize=(10, 6))
data_exp = np.reshape(expvals_exp, (points, len(steps)))
colors = ['#0f62fe', '#be95ff', '#ff7eb6']
for i in range(points):
    plt.plot(steps, data_exp[i],
             label=f'theta={i*max_angle/(points-1)} (full CX)', 
             linestyle='dashed', color=colors[i], lw=2)

plt.xlabel('Trotter steps', fontsize=16)
plt.ylabel('Average magnetization', fontsize=16)
plt.xticks(fontsize=13)
plt.yticks(fontsize=13)
plt.xticks(rotation=45)
handles, labels = plt.gca().get_legend_handles_labels()
plt.legend(handles, labels, loc='lower right', 
bbox_to_anchor=(1.0, -0.01), shadow=True, ncol=3)
plt.show()
\end{pythonblock}

\subsection{Retargetable transpilation}

A key feature of the Qiskit transpiler is that it is retargetable to a variety of gatesets, which can include non-standard and heterogeneous gates. Continuing with the same example, suppose now that our quantum computer has a new gate available in its ISA, for example a partial $ZX$ rotation. In our considered hardware this is achievable, for example, by applying the cross-resonance interaction for a shorter duration of time~\cite{earnest2021pulse}. We do not detail the calibration of such gates here, but this is also possible through Qiskit~\cite{alexander2020qiskit}. For our purpose, all we need to do is to enhance the target's ISA to include information about this new gate, including its definition and the fact that it has shorter duration and error compared to a full CX. An example modified target will look as follows.

\begin{pythonblock}
# step 2: modified transpile 

from qiskit.transpiler import InstructionProperties
from qiskit.circuit.library import RZXGate

rzx_props = {}
for edge in target['ecr'].keys():
    ecr_dur = target['ecr'][edge].duration
    ecr_err = target['ecr'][edge].error
    op = RZXGate(np.pi/8)
    rzx_err = 0.25 * ecr_dur
    rzx_dur = 0.25 * ecr_err
    rzx_props[edge] = InstructionProperties(duration=rzx_dur,
                                            error=rzx_err)
target.add_instruction(op, rzx_props, name='rzx')
\end{pythonblock}

Qiskit is now able to utilize this overcomplete gateset to shorten the total duration of circuits, boosting output fidelity. It is not even necessary for the gate to be calibrated on all edges---the transpiler can use it opportunistically where possible. To take advantage of these newly calibrated gates, we re-transpile our circuits using the same procedure as before but now with the new target definition. In this particular example, Qiskit uses the method from~\cite{peterson2022optimal} to synthesize $ZZ(\theta)$ interactions occurring in each Trotter layer to $ZX(\phi)$ interactions available on hardware, which looks as follows for any $\phi \leq \theta/2$.

\vspace{3mm}

\begin{tikzpicture}[scale=1, font=\footnotesize]
   \begin{yquantgroup}
      \registers{
         qubit {} q[2];
      }
      % \circuit{
      %    zz (q[1], q[0]);
      % }
      % \equals
      % \circuit{
      %    cnot q[1] | q[0];
      %    box {$Z_{\theta}$} q[1];
      %    cnot q[1] | q[0];
      % }
      % \equals
      \circuit{
         [x radius=.9cm] box {$X_\frac{\pi}{2}Z_{\lambda_1}X_\frac{\pi}{2}$} q[0];
         [x radius=.8cm] box {$Z_\frac{\pi}{2}X_{\frac{-\pi}{2}}Z_\frac{\pi}{2}$} q[1];
         [x radius=.2cm] box {$ZX_\phi$} (q[0] , q[1]);
         [x radius=.9cm] box {$X_\frac{\pi}{2}Z_{\lambda_2}X_\frac{\pi}{2}$} q[0];
         [x radius=.2cm]box {$ZX_\phi$} (q[0] , q[1]);
         [x radius=.9cm] box {$X_\frac{\pi}{2}Z_{\lambda_1}X_\frac{\pi}{2}$} q[0];
         [x radius=.9cm] box {$Z_\frac{\pi}{2}X_{\frac{-\pi}{2}}Z_\frac{\pi}{2}$} q[1];
         z q[0];
      }
   \end{yquantgroup}
\end{tikzpicture}

\vspace{3mm}

In our example, we see that transpiling to this new gateset allows us to achieve significantly reduced circuit durations, which correspondingly translate to improved Ising model simulations (see bottom left of Figure~\ref{fig:ising}(b) and left of Figure~\ref{fig:ising}(c)).

\subsection{Dynamic circuit adaptation}

In the previous section we simulated a lattice of spins which resembled our available hardware. Now suppose that we wish to simulate a honeycomb (hexagonal) lattice. In contrast to the heavy-hex lattice where the average node has degree 2.4 (the nodes alternate between degrees 2 and 3), nodes in the honeycomb lattice have degree 3 which makes the system harder to simulate classically. 

The straightforward way to perform this experiment is to use Qiskit's transpiler to map the problem to the hardware, which will involve a series of SWAP operations to bring interacting qubits close to each other as the circuit progresses. Here we use dynamic circuits to solve the problem in shorter depth. This illustrates the power of mixing quantum and (real-time) classical computation within a circuit in Qiskit. In this example, our dynamic circuit will consist of mid-circuit measurements, real-time classical computation on measured bits, classical control flow depending on the result of classical computations, and ancilla reset. This circuit is also similar to protocols in fault-tolerant quantum computing where syndromes are measured from ancilla and corrections are applied on data. Our method and results are depicted on the right hand side of Figure~\ref{fig:ising}.

Below we build a dynamic circuit for this task and use it to measure the average magnetization as before. 

\begin{pythonblock}
from qiskit.circuit import ClassicalRegister, QuantumRegister
from qiskit.circuit.classical import expr
from qiskit.transpiler import CouplingMap

# define lattice
hex_rows = 1
hex_cols = 1
hex_cmap = CouplingMap.from_hexagonal_lattice(
    hex_rows, hex_cols, bidirectional=False)
data = list(hex_cmap.physical_qubits)

# step 1: build dynamic circuits and define observable
heavyhex_cmap = CouplingMap()
for d in data:
    heavyhex_cmap.add_physical_qubit(d)
a = len(data)
for edge in hex_cmap.get_edges():
    heavyhex_cmap.add_physical_qubit(a)
    heavyhex_cmap.add_edge(edge[0], a)
    heavyhex_cmap.add_edge(edge[1], a)
    a += 1
ancilla = list(range(len(data), a))
qubits = data + ancilla
graph = heavyhex_cmap.graph.to_undirected(multigraph=False)
edge_colors = rx.graph_misra_gries_edge_color(graph)
layer_edges = {color: [] for color in edge_colors.values()}
for edge_index, color in edge_colors.items():
    layer_edges[color].append(graph.edge_list()[edge_index])

circuits_hex = []
steps = range(10)
zz_angle = np.pi/8
theta = Parameter('theta')
for step in steps:
    qr = QuantumRegister(len(qubits), 'qr')
    cr = ClassicalRegister(len(ancilla), 'cr')
    circuit =  QuantumCircuit(qr, cr)
    for d in data:
        circuit.rx(theta, d)
    # trotter steps expressed as a circuit loop
    with circuit.for_loop(range(step)):
        circuit.barrier()
        # computing parities into ancilla
        for same_color_edges in layer_edges.values():
            for e in same_color_edges:
                circuit.cx(e[0], e[1])
        circuit.barrier()
        # applying rotation and measuring out ancilla
        for i, a in enumerate(ancilla):
            circuit.rz(zz_angle, a)
            circuit.h(a)
            circuit.measure(a, i)
        d2ros = {}   # each data contributes to 2 neighboring ROs
        a2ro = {}    # each ancilla is measured into 1 readout
        for a in ancilla:
            a2ro[a] = cr[ancilla.index(a)]
        for d in data:
            ros = [a2ro[a] for a in heavyhex_cmap.neighbors(d)]
            d2ros[d] = ros
        # determine if a data qubit should be Pauli-corrected
        # XOR its neighboring readouts, if True apply Z to it
        for d in data:
            ros = d2ros[d]
            parity = ros[0]
            for ro in ros[1:]:
                parity = expr.bit_xor(parity, ro)
            with circuit.if_test(expr.equal(parity, True)):
                circuit.z(d)
        # determine if an ancilla should be reset
        # if its readout is True, then flip it
        for a in ancilla:
            with circuit.if_test(expr.equal(a2ro[a], True)):
                circuit.x(a)
        circuit.barrier()
        # Ising kick
        for d in data:
            circuit.rx(theta, d)
    circuits_hex.append(circuit)

obs_hex = SparsePauliOp.from_sparse_list(
    [("Z", [i], 1/len(data)) for i in data], 
    num_qubits=len(qubits))
\end{pythonblock}

In our circuit, we effect the $ZZ$ rotations by using ancilla qubits, mid-circuit measurements, and feed-forward. To understand this, note that $ZZ$ rotations apply a phase to those computational basis states whose parity is odd ($\ket{01}$ and $\ket{10}$). Therefore we need to compute the parity of the qubits on which we wish to apply the $ZZ$ rotation. This is essentially what the $CX$ gate does. However, it is also possible to compute the parity into a third (ancilla) qubit, and apply the $ZZ$ rotation as a single-qubit $Z$ rotation acting on the ancilla. Now we can measure out the ancilla in the $X$ basis. When a $0$ outcome is observed, we have in fact correctly applied a $ZZ(\theta)$ rotation to our data qubits. If, however, a $1$ outcome is observed, it means a $ZZ(\theta+\pi)$ rotation has instead been applied. We can ``fix'' this by applying a Pauli-$Z$ correction to the data qubits whenever the ancilla is measured as $1$. This is equivalent to the following circuit identity.

\vspace{3mm}
\begin{tikzpicture}[scale=1, font=\footnotesize]
   \begin{yquantgroup}
      \registers{
         qubit {} q[3];
      }
      \circuit{
         discard q[1];
         zz (q[2], q[0]);
      }
      \equals
      \circuit{
         discard q[1];
         cnot q[2] | q[0];
         box {$Z_{\theta}$} q[2];
         cnot q[2] | q[0];
      }
      \equals
      \circuit{
         init {$\ket0$} q[1];
         cnot q[1] | q[0];
         cnot q[1] | q[2];
         box {$Z_{\theta}$} q[1];
         h q[1];
         [x radius=.25cm] measure q[1];
         z q[0], q[2] | q[1];
         % x q[1] | q[1];
         % init {$\ket0$} q[1];
         discard q[1];

      }
   \end{yquantgroup}
\end{tikzpicture}

\vspace{3mm}

Even though this approach seems more costly as it uses the same number of CX gates, as well as an extra ancilla and measurement, the advantage of the measurement-based circuit becomes evident when we consider multiple $ZZ$ interactions --- the CX layers can all parallelize, and measurements can all occur simultaneously. This is due to the fact that all $ZZ$ interactions commute and so it is possible to perform the computation in measurement-depth of 1. Furthermore, these circuits embed perfectly into the heavy-hex lattice of the quantum computer: all data qubits reside on the degree-3 sites of the lattice, which forms a hexagonal lattice. Every pair of data qubits shares an ancilla qubit, which resides on the degree-2 sites. See Figure~\ref{fig:ising}(b) (right) for the approach as well as the savings in circuit depth compared to the mapping and routing approach.

We entangle all data with their corresponding ancilla using parallel CX layers. We then perform a layer of $Z$ rotations, followed by a layer of ancilla measurement in the $X$ basis, followed by a conditional Pauli-correction layer. Note that the ancilla must be reset for the next Trotter step, but we can perform this at no extra cost by performing conditional-$X$ operations on the ancilla at the same time as we are correcting the data with conditional-$Z$ operations. For simplicity we study a small instance of this problem, corresponding to a single 6-qubit cell of the honeycomb lattice (leading to a 12-qubit circuit including ancilla). We can choose which hardware qubits our circuits maps to.

\begin{pythonblock}
# step 2: transpile so circuit is laid out on physical qubits
layout = [104, 122, 124, 106, 108, 126,
          111, 123, 107, 112, 105, 125]
pm = generate_preset_pass_manager(
    target=target, 
    optimization_level=3, 
    initial_layout=layout
)
circuits_isa = pm.run(circuits_hex)
obs_isa = o.apply_layout(circuits_isa[0].layout)
\end{pythonblock}

The execution via the estimator and analysis of results is exactly as before. The results are shown in Figure~\ref{fig:ising}(c) (right). We can see a slight improvement with the use of dynamics circuits compared to the swap-based approach. 

Note that in all of these experiments we have not performed error mitigation methods that incur a large sampling overhead, although these methods are expected to extend the computational reach of circuits further~\cite{van2023probabilistic}. However, we have performed error suppression where possible to counter the effect of known errors. For example the use of dynamical decoupling in the previous section was a simple way to refocus coherent noise and suppress incoherent noise to first order. This becomes even more important with dynamic circuits due to the long periods of idleness on some qubits while measurement and feed-forward is occurring in the circuit. We use noise suppression methods tailored to dynamic circuits in general~\cite{seif2024suppressing, vazquez2024scaling} and to our specific circuits. Intuitively, we cancel the effect of coherent $ZZ$ accumulation during these idle periods by effecting a slightly different $ZZ$ rotation in our circuits.  The experimental details of our error suppression procedure are detailed in Appendix~\ref{sec:error-suppression}.
%There are other approaches to using ancilla to emulate a honeycomb lattice with a heavy-hex lattice that use neither swaps nor dynamic circuits, but our goal here was to illustrate the use of dynamic circuits which in general can provide depth reductions

% old examples
% \input{example-hubbard}
% \input{example-hs}

\section{Qiskit Ecosystem}\label{sec:ecosystem}

Qiskit has served as a catalyst towards the development of a large ecosystem of open-source software around quantum computing. These packages have been built on top of Qiskit and extend the functionality that Qiskit itself provides~\cite{ecosystem}. Some particular examples include high-performance simulation of quantum circuits~\cite{qiskit-aer, qiskit-qualcs}, tools for calibration or characterization of quantum hardware and simulation of open quantum systems~\cite{qiskit-experiments, c3, qiskit-dynamics}, quantum algorithms and applications such as in chemistry and machine learning~\cite{qiskit-nature,hanruiwang2022quantumnas}, and plugin transpiler passes~\cite{dsm-swap, sat-synthesis, toqm, bip}.

Likewise, Qiskit has been used in many scientific research projects, many of which provide accompanying Qiskit-compatible codes for use by other researchers. This includes novel methods for circuit transformation and resource optimization~\cite{murali2019noise, iten2022exact, lao20222qan, mckinney2023mirage, dangwal2023varsaw, zhang2023oneq, hua2023caqr, tang2021cutqc, weaving2023stabilizer, knitting}, using pulse and timing capabilities for error suppression or enablement of new gates~\cite{murali2020software, smith2021error, earnest2021pulse, mundada2023experimental, superstaq, wei2023native}, or high-level language design and formal compiler verification~\cite{seidel2022qrisp,tao2022giallar}.

As we emphasized in previous sections, Qiskit's flexible circuit model and retargetable transpiler have allowed its scope to be expanded. It has been used to represent computations on qudits and Bosonic modes, to represent measurement-based and Pauli-based quantum computations, as well as to represent incoherent quantum noise channels~\cite{stavenger2022c2qa, litteken2023qompress, kashif2022qiskit, peres2023quantum, horii2023efficient}. It has also been used to target a variety of technological platforms such superconducting, trapped ion and cold atom quantum computers~\cite{qiskit-ibm-runtime, qiskit-ionq, qiskit-cold-atom}.

More broadly, the OpenQASM language is now widely supported among other quantum tools, which serves to connect Qiskit to an even wider ecosystem of quantum software. Some of these tools are explicitly built around OpenQASM~\cite{staq, oqpy, autoqasm, qasmtrans}, while others support serialization to and from it~\cite{pennylane, projectq, braket, cirq, scaffold, pyzx, tket, quipper, quilc}.

% \section{Related Work}\label{sec:related}

% There have been many software platforms related to quantum information science~\cite{strawberryfields, staq, scaffcc, projectq, quipper, qsharp, cirq, forest}

% Qiskit remains the most widely used, and the most comprehensive.

\section{Conclusions}\label{sec:conclusion}

With rapid progress in experimental and theoretical quantum computing, the field is at the cusp of surpassing classical computation for useful tasks. The road ahead is likely to rely heavily on the co-design of software and hardware, making it necessary to explore different choices of architectures, compilers, error correcting codes, decoders, error suppression and mitigation strategies, and algorithms and applications. Quantum software development kits such as Qiskit can be indispensable tools for enabling this research and development. 

In this paper we have discussed Qiskit's software architecture, its key components, and some examples of building workflows to explore quantum algorithms on today's hardware. With the release of version 1.0, Qiskit has reached a stage of maturity, stability and performance that enable it to be used as part of utility-scale quantum computational workflows.

The future holds many exciting directions. Successful quantum error correction will push logical error rates down by multiple orders of magnitude, which correspondingly mean orders of magnitude larger circuits will routinely be built and executed. Qiskit's performance will correspondingly improve to carry forward into the fault-tolerant era. Qiskit's ability to reason about structured and hierarchical circuits will also improve to enable better representation of fault-tolerant protocols and computations on modular architectures. 
\section{Acknowledgments}\label{sec:ack}
We thank Qiskit's many individual contributors who have played an instrumental role in making the project successful~\cite{contributors}. In particular we are grateful for significant contributions from Thomas Alexander, Luciano Bello, Sebastian Brandhofer, Lauren Capelluto, Daniel Egger, Ismael Faro, Shelly Garion, Juan Gomez, Ikko Hamamura, Kevin Hartman, Ian Hincks, Takashi Imamichi, Alexander Ivrii, Naoki Kanazawa, John Lapeyre, Manoel Marques, Dmitri Maslov, David McKay, Antonio Mezzacapo, Diego Moreda, Ed Navarro, Elena Pe{\~n}a Tapia, Eric Peterson, Kevin Sung, Soolu Thomas, Erick Winston, Stefan Woerner and Stephen Wood. We thank Diego Rist{\`e}, Alireza Seif and Maika Takita for help with the dynamic circuit experiment in this manuscript. The development of Qiskit was partially supported by the U.S. Department of Energy, Office of Science, National Quantum Information Science Research Centers, Co-design Center for Quantum Advantage (C2QA) under contract number DE-SC0012704.

\bibliographystyle{plainnat} 
\bibliography{refs}

\begin{thebibliography}{104}
\providecommand{\natexlab}[1]{#1}
\providecommand{\url}[1]{\texttt{#1}}
\expandafter\ifx\csname urlstyle\endcsname\relax
  \providecommand{\doi}[1]{doi: #1}\else
  \providecommand{\doi}{doi: \begingroup \urlstyle{rm}\Url}\fi

\bibitem[asv()]{asv}
{Airspeed velocity of an unladen Qiskit}.
\newblock URL \url{https://qiskit.github.io/qiskit/}.

\bibitem[con()]{contributors}
{Qiskit contributors}.
\newblock URL \url{https://github.com/Qiskit/qiskit/graphs/contributors}.

\bibitem[git()]{github}
{Qiskit on Github}.
\newblock URL \url{https://github.com/Qiskit/qiskit}.

\bibitem[pyp()]{pypi}
{Qiskit on the Python Package Index (PyPI)}.
\newblock URL \url{https://pypi.org/project/qiskit/}.

\bibitem[qis({\natexlab{a}})]{qiskit-cold-atom}
{Qiskit Cold Atom}, {\natexlab{a}}.
\newblock URL \url{https://github.com/qiskit-community/qiskit-cold-atom}.

\bibitem[qis({\natexlab{b}})]{qiskit-ibm-runtime}
{Qiskit IBM Runtime}, {\natexlab{b}}.
\newblock URL \url{https://github.com/Qiskit/qiskit-ibm-runtime}.

\bibitem[qis({\natexlab{c}})]{qiskit-ionq}
{Qiskit IonQ}, {\natexlab{c}}.
\newblock URL \url{https://github.com/Qiskit/qiskit-ionq}.

\bibitem[qis({\natexlab{d}})]{qiskit-rigetti}
{Qiskit Rigetti}, {\natexlab{d}}.
\newblock URL \url{https://github.com/Qiskit/qiskit-rigetti}.

\bibitem[qui()]{quirk}
Quirk.
\newblock URL \url{https://algassert.com/quirk}.

\bibitem[uni()]{unitaryfund}
{The State of quantum open source software 2023: survey results}.
\newblock URL \url{https://unitary.fund/posts/2023_survey_results}.

\bibitem[eco(2024)]{ecosystem}
{Qiskit Ecosystem}, 2024.
\newblock URL \url{https://qiskit.github.io/ecosystem/}.

\bibitem[kni(2024)]{knitting}
{Circuit Knitting Toolbox}, 2024.
\newblock URL \url{https://github.com/Qiskit-Extensions/circuit-knitting-toolbox}.

\bibitem[sat(2024)]{sat-synthesis}
{Qiskit SAT Synthesis}, 2024.
\newblock URL \url{https://github.com/qiskit-community/qiskit-sat-synthesis}.

\bibitem[Aaronson and Gottesman(2004)]{aaronson2004improved}
Scott Aaronson and Daniel Gottesman.
\newblock Improved simulation of stabilizer circuits.
\newblock \emph{Physical Review A}, 70\penalty0 (5):\penalty0 052328, 2004.

\bibitem[Alexander et~al.(2020)Alexander, Kanazawa, Egger, Capelluto, Wood, Javadi-Abhari, and McKay]{alexander2020qiskit}
Thomas Alexander, Naoki Kanazawa, Daniel~J Egger, Lauren Capelluto, Christopher~J Wood, Ali Javadi-Abhari, and David~C McKay.
\newblock Qiskit pulse: programming quantum computers through the cloud with pulses.
\newblock \emph{Quantum Science and Technology}, 5\penalty0 (4):\penalty0 044006, 2020.

\bibitem[Alexeev et~al.(2024)Alexeev, Amsler, Barroca, Bassini, Battelle, Camps, Casanova, jai Choi, Chong, Chung, et~al.]{alexeev2024quantum}
Yuri Alexeev, Maximilian Amsler, Marco~Antonio Barroca, Sanzio Bassini, Torey Battelle, Daan Camps, David Casanova, Young jai Choi, Frederic~T Chong, Charles Chung, et~al.
\newblock Quantum-centric supercomputing for materials science: A perspective on challenges and future directions.
\newblock \emph{Future Generation Computer Systems}, 2024.

\bibitem[{Amazon Web Services}(2020)]{braket}
{Amazon Web Services}.
\newblock {Amazon Braket}, 2020.
\newblock URL \url{https://aws.amazon.com/braket/}.

\bibitem[{Amazon Web Services}(2024)]{autoqasm}
{Amazon Web Services}.
\newblock {AutoQASM}, 2024.
\newblock URL \url{https://github.com/amazon-braket/amazon-braket-sdk-python/tree/feature/autoqasm/src/braket/experimental/autoqasm}.

\bibitem[Amy and Gheorghiu(2020)]{staq}
Matthew Amy and Vlad Gheorghiu.
\newblock staq---a full-stack quantum processing toolkit.
\newblock \emph{Quantum Science and Technology}, 5\penalty0 (3):\penalty0 034016, 2020.
\newblock URL \url{https://github.com/softwareQinc/staq}.

\bibitem[Amy et~al.(2014)Amy, Maslov, and Mosca]{amy2014polynomial}
Matthew Amy, Dmitri Maslov, and Michele Mosca.
\newblock Polynomial-time {T}-depth optimization of {Clifford}+{T} circuits via matroid partitioning.
\newblock \emph{IEEE Transactions on Computer-Aided Design of Integrated Circuits and Systems}, 33\penalty0 (10):\penalty0 1476--1489, 2014.

\bibitem[B\"{a}umer et~al.(2023)B\"{a}umer, Tripathi, Wang, Rall, Chen, Majumdar, Seif, and Minev]{baumer:2023}
Elisa B\"{a}umer, Vinay Tripathi, Derek~S. Wang, Patrick Rall, Edward~H. Chen, Swarnadeep Majumdar, Alireza Seif, and Zlatko~K Minev.
\newblock {E}fficient {L}ong-{R}ange {E}ntanglement using {D}ynamic {C}ircuits.
\newblock \emph{arXiv preprint arXiv:2308.13065}, 2023.
\newblock \doi{10.48550/arXiv.2308.13065}.

\bibitem[Bennett et~al.(1996{\natexlab{a}})Bennett, Brassard, Popescu, Schumacher, Smolin, and Wootters]{bennett1996purification}
Charles~H Bennett, Gilles Brassard, Sandu Popescu, Benjamin Schumacher, John~A Smolin, and William~K Wootters.
\newblock Purification of noisy entanglement and faithful teleportation via noisy channels.
\newblock \emph{Physical review letters}, 76\penalty0 (5):\penalty0 722, 1996{\natexlab{a}}.

\bibitem[Bennett et~al.(1996{\natexlab{b}})Bennett, DiVincenzo, Smolin, and Wootters]{bennett1996mixed}
Charles~H Bennett, David~P DiVincenzo, John~A Smolin, and William~K Wootters.
\newblock Mixed-state entanglement and quantum error correction.
\newblock \emph{Physical Review A}, 54\penalty0 (5):\penalty0 3824, 1996{\natexlab{b}}.

\bibitem[Bergholm et~al.(2018)Bergholm, Izaac, Schuld, Gogolin, Ahmed, Ajith, Alam, Alonso-Linaje, AkashNarayanan, Asadi, et~al.]{pennylane}
Ville Bergholm, Josh Izaac, Maria Schuld, Christian Gogolin, Shahnawaz Ahmed, Vishnu Ajith, M~Sohaib Alam, Guillermo Alonso-Linaje, B~AkashNarayanan, Ali Asadi, et~al.
\newblock Pennylane: Automatic differentiation of hybrid quantum-classical computations.
\newblock \emph{arXiv preprint arXiv:1811.04968}, 2018.
\newblock URL \url{https://github.com/PennyLaneAI/pennylane}.

\bibitem[Beverland et~al.(2022)Beverland, Murali, Troyer, Svore, Hoefler, Kliuchnikov, Low, Soeken, Sundaram, and Vaschillo]{beverland2022assessing}
Michael~E Beverland, Prakash Murali, Matthias Troyer, Krysta~M Svore, Torsten Hoefler, Vadym Kliuchnikov, Guang~Hao Low, Mathias Soeken, Aarthi Sundaram, and Alexander Vaschillo.
\newblock Assessing requirements to scale to practical quantum advantage.
\newblock \emph{arXiv preprint arXiv:2211.07629}, 2022.

\bibitem[Bravyi and Kitaev(2005)]{bravyi2005universal}
Sergey Bravyi and Alexei Kitaev.
\newblock Universal quantum computation with ideal {Clifford} gates and noisy ancillas.
\newblock \emph{Physical Review A}, 71\penalty0 (2):\penalty0 022316, 2005.

\bibitem[Bravyi et~al.(2020)Bravyi, Gosset, Koenig, and Tomamichel]{bravyi2020quantum}
Sergey Bravyi, David Gosset, Robert Koenig, and Marco Tomamichel.
\newblock Quantum advantage with noisy shallow circuits.
\newblock \emph{Nature Physics}, 16\penalty0 (10):\penalty0 1040--1045, 2020.

\bibitem[Campbell et~al.(2023)Campbell, Chong, Dahl, Frederick, Goiporia, Gokhale, Hall, Issa, Jones, Lee, et~al.]{superstaq}
Colin Campbell, Frederic~T Chong, Denny Dahl, Paige Frederick, Palash Goiporia, Pranav Gokhale, Benjamin Hall, Salahedeen Issa, Eric Jones, Stephanie Lee, et~al.
\newblock Superstaq: Deep optimization of quantum programs.
\newblock In \emph{2023 IEEE International Conference on Quantum Computing and Engineering (QCE)}, volume~1, pages 1020--1032. IEEE, 2023.
\newblock URL \url{https://github.com/Infleqtion/client-superstaq}.

\bibitem[Cerezo et~al.(2021)Cerezo, Arrasmith, Babbush, Benjamin, Endo, Fujii, McClean, Mitarai, Yuan, Cincio, et~al.]{cerezo2021variational}
Marco Cerezo, Andrew Arrasmith, Ryan Babbush, Simon~C Benjamin, Suguru Endo, Keisuke Fujii, Jarrod~R McClean, Kosuke Mitarai, Xiao Yuan, Lukasz Cincio, et~al.
\newblock Variational quantum algorithms.
\newblock \emph{Nature Reviews Physics}, 3\penalty0 (9):\penalty0 625--644, 2021.

\bibitem[Chen et~al.(2023)Chen, Zhu, Verresen, Seif, B\"{a}umer, Layden, Tantivasadakarn, Zhu, Sheldon, Vishwanath, Trebst, and Kandala]{chen:2023}
Edward~H. Chen, Guo-Yi Zhu, Ruben Verresen, Alireza Seif, Elisa B\"{a}umer, David Layden, Nathanan Tantivasadakarn, Guanyu Zhu, Sarah Sheldon, Ashvin Vishwanath, Simon Trebst, and Abhinav Kandala.
\newblock {R}ealizing the {N}ishimori transition across the error threshold for constant-depth quantum circuits.
\newblock \emph{arXiv preprint arXiv:2309.02863}, 2023.
\newblock \doi{10.48550/arXiv.2309.02863}.

\bibitem[Cross et~al.(2022)Cross, Javadi-Abhari, Alexander, De~Beaudrap, Bishop, Heidel, Ryan, Sivarajah, Smolin, Gambetta, et~al.]{cross2022openqasm}
Andrew Cross, Ali Javadi-Abhari, Thomas Alexander, Niel De~Beaudrap, Lev~S Bishop, Steven Heidel, Colm~A Ryan, Prasahnt Sivarajah, John Smolin, Jay~M Gambetta, et~al.
\newblock {OpenQASM} 3: A broader and deeper quantum assembly language.
\newblock \emph{ACM Transactions on Quantum Computing}, 3\penalty0 (3):\penalty0 1--50, 2022.

\bibitem[Dangwal et~al.(2023)Dangwal, Ravi, Das, Smith, Baker, and Chong]{dangwal2023varsaw}
Siddharth Dangwal, Gokul~Subramanian Ravi, Poulami Das, Kaitlin~N Smith, Jonathan~Mark Baker, and Frederic~T Chong.
\newblock {VarSaw}: Application-tailored measurement error mitigation for variational quantum algorithms.
\newblock In \emph{Proceedings of the 28th ACM International Conference on Architectural Support for Programming Languages and Operating Systems, Volume 4}, pages 362--377, 2023.
\newblock URL \url{https://github.com/siddharthdangwal/VarSaw}.

\bibitem[Developers(2023)]{cirq}
Cirq Developers.
\newblock Cirq, July 2023.
\newblock URL \url{https://github.com/quantumlib/Cirq}.

\bibitem[developers and contributors(2023)]{qiskit-nature}
The Qiskit~Nature developers and contributors.
\newblock Qiskit nature, April 2023.
\newblock URL \url{https://doi.org/10.5281/zenodo.8161252}.

\bibitem[Doi et~al.(2023)Doi, Horii, and Wood]{horii2023efficient}
Jun Doi, Hiroshi Horii, and Christopher~J. Wood.
\newblock Efficient techniques to {GPU} accelerations of multi-shot quantum computing simulations.
\newblock \emph{arXiv preprint arXiv:2308.03399}, 2023.

\bibitem[Earnest et~al.(2021)Earnest, Tornow, and Egger]{earnest2021pulse}
Nathan Earnest, Caroline Tornow, and Daniel~J Egger.
\newblock Pulse-efficient circuit transpilation for quantum applications on cross-resonance-based hardware.
\newblock \emph{Physical Review Research}, 3\penalty0 (4):\penalty0 043088, 2021.

\bibitem[Elben et~al.(2023)Elben, Flammia, Huang, Kueng, Preskill, Vermersch, and Zoller]{elben2023randomized}
Andreas Elben, Steven~T Flammia, Hsin-Yuan Huang, Richard Kueng, John Preskill, Beno{\^\i}t Vermersch, and Peter Zoller.
\newblock The randomized measurement toolbox.
\newblock \emph{Nature Reviews Physics}, 5\penalty0 (1):\penalty0 9--24, 2023.

\bibitem[Faro et~al.(2023)Faro, Sitdikov, Vali{\~n}as, Fernandez, Codella, and Glick]{faro2023middleware}
Ismael Faro, Iskandar Sitdikov, David~Garcia Vali{\~n}as, Francisco Jose~Martin Fernandez, Christopher Codella, and Jennifer Glick.
\newblock Middleware for quantum: An orchestration of hybrid quantum-classical systems.
\newblock In \emph{2023 IEEE International Conference on Quantum Software (QSW)}, pages 1--8. IEEE, 2023.

\bibitem[Farrell et~al.(2024{\natexlab{a}})Farrell, Illa, Ciavarella, and Savage]{farrell2023scalable}
Roland~C Farrell, Marc Illa, Anthony~N Ciavarella, and Martin~J Savage.
\newblock Scalable circuits for preparing ground states on digital quantum computers: The schwinger model vacuum on 100 qubits.
\newblock \emph{PRX Quantum}, 5:\penalty0 020315, Apr 2024{\natexlab{a}}.
\newblock \doi{10.1103/PRXQuantum.5.020315}.
\newblock URL \url{http://dx.doi.org/10.1103/PRXQuantum.5.020315}.

\bibitem[Farrell et~al.(2024{\natexlab{b}})Farrell, Illa, Ciavarella, and Savage]{farrell:2024}
Roland~C Farrell, Marc Illa, Anthony~N Ciavarella, and Martin~J Savage.
\newblock {Q}uantum {S}imulations of {H}adron {D}ynamics in the {S}chwinger {M}odel using 112 {Q}ubits.
\newblock \emph{arXiv preprint arXiv:2401.08044}, 2024{\natexlab{b}}.
\newblock \doi{10.48550/arXiv.2401.08044}.

\bibitem[Green et~al.(2013)Green, Lumsdaine, Ross, Selinger, and Valiron]{quipper}
Alexander~S Green, Peter~LeFanu Lumsdaine, Neil~J Ross, Peter Selinger, and Beno{\^\i}t Valiron.
\newblock Quipper: a scalable quantum programming language.
\newblock In \emph{Proceedings of the 34th ACM SIGPLAN conference on Programming language design and implementation}, pages 333--342, 2013.
\newblock URL \url{https://www.mathstat.dal.ca/~selinger/quipper/}.

\bibitem[Gupta et~al.(2023)Gupta, Sundaresan, Alexander, Wood, Merkel, Healy, Hillenbrand, Jochym-O'Connor, Wootton, Yoder, Cross, Takita, and Brown]{gupta2023encoding}
Riddhi~S Gupta, Neereja Sundaresan, Thomas Alexander, Christopher~J Wood, Seth~T Merkel, Michael~B Healy, Marius Hillenbrand, Tomas Jochym-O'Connor, James~R Wootton, Theodore~J Yoder, Andrew~W Cross, Maika Takita, and Benjamin~J Brown.
\newblock Encoding a magic state with beyond break-even fidelity.
\newblock \emph{arXiv preprint arXiv:2305.13581}, 2023.

\bibitem[Horii et~al.(2023)Horii, Wood, et~al.]{qiskit-aer}
Hiroshi Horii, Christopher Wood, et~al.
\newblock Efficient techniques to gpu accelerations of multi-shot quantum computing simulations.
\newblock \emph{arXiv preprint arXiv:2308.03399}, 2023.

\bibitem[Hua et~al.(2023{\natexlab{a}})Hua, Jin, Chen, Vittal, Krsulich, Bishop, Lapeyre, Javadi-Abhari, and Zhang]{hua2023caqr}
Fei Hua, Yuwei Jin, Yanhao Chen, Suhas Vittal, Kevin Krsulich, Lev~S Bishop, John Lapeyre, Ali Javadi-Abhari, and Eddy~Z Zhang.
\newblock {CaQR}: A compiler-assisted approach for qubit reuse through dynamic circuit.
\newblock In \emph{Proceedings of the 28th ACM International Conference on Architectural Support for Programming Languages and Operating Systems, Volume 3}, pages 59--71, 2023{\natexlab{a}}.
\newblock URL \url{https://github.com/ruadapt/CaQR}.

\bibitem[Hua et~al.(2023{\natexlab{b}})Hua, Wang, Li, Peng, Liu, Zheng, Stein, Ding, Zhang, Humble, et~al.]{qasmtrans}
Fei Hua, Meng Wang, Gushu Li, Bo~Peng, Chenxu Liu, Muqing Zheng, Samuel Stein, Yufei Ding, Eddy~Z Zhang, Travis~S Humble, et~al.
\newblock Qasmtrans: A qasm based quantum transpiler framework for nisq devices.
\newblock \emph{arXiv preprint arXiv:2308.07581}, 2023{\natexlab{b}}.
\newblock URL \url{https://github.com/pnnl/qasmtrans}.

\bibitem[Huang et~al.(2020)Huang, Kueng, and Preskill]{huang2020predicting}
Hsin-Yuan Huang, Richard Kueng, and John Preskill.
\newblock Predicting many properties of a quantum system from very few measurements.
\newblock \emph{Nature Physics}, 16\penalty0 (10):\penalty0 1050--1057, 2020.

\bibitem[Iten et~al.(2022)Iten, Moyard, Metger, Sutter, and Woerner]{iten2022exact}
Raban Iten, Romain Moyard, Tony Metger, David Sutter, and Stefan Woerner.
\newblock Exact and practical pattern matching for quantum circuit optimization.
\newblock \emph{ACM Transactions on Quantum Computing}, 3\penalty0 (1):\penalty0 1--41, 2022.

\bibitem[Javadi-Abhari et~al.(2015)Javadi-Abhari, Patil, Kudrow, Heckey, Lvov, Chong, and Martonosi]{scaffold}
Ali Javadi-Abhari, Shruti Patil, Daniel Kudrow, Jeff Heckey, Alexey Lvov, Frederic~T Chong, and Margaret Martonosi.
\newblock Scaffcc: Scalable compilation and analysis of quantum programs.
\newblock \emph{Parallel Computing}, 45:\penalty0 2--17, 2015.
\newblock URL \url{https://github.com/epiqc/ScaffCC}.

\bibitem[Kanazawa et~al.(2023)Kanazawa, Egger, Ben-Haim, Zhang, Shanks, Aleksandrowicz, and Wood]{qiskit-experiments}
Naoki Kanazawa, Daniel~J Egger, Yael Ben-Haim, Helena Zhang, William~E Shanks, Gadi Aleksandrowicz, and Christopher~J Wood.
\newblock Qiskit experiments: A python package to characterize and calibrate quantum computers.
\newblock \emph{Journal of Open Source Software}, 8\penalty0 (84):\penalty0 5329, 2023.

\bibitem[Kashif and Al-Kuwari(2022)]{kashif2022qiskit}
Muhammad Kashif and Saif Al-Kuwari.
\newblock Qiskit as a simulation platform for measurement-based quantum computation.
\newblock In \emph{2022 IEEE 19th International Conference on Software Architecture Companion (ICSA-C)}, pages 152--159. IEEE, 2022.

\bibitem[Kim et~al.(2023)Kim, Eddins, Anand, Wei, Van Den~Berg, Rosenblatt, Nayfeh, Wu, Zaletel, Temme, and Kandala]{kim2023evidence}
Youngseok Kim, Andrew Eddins, Sajant Anand, Ken~Xuan Wei, Ewout Van Den~Berg, Sami Rosenblatt, Hasan Nayfeh, Yantao Wu, Michael Zaletel, Kristan Temme, and Abhinav Kandala.
\newblock Evidence for the utility of quantum computing before fault tolerance.
\newblock \emph{Nature}, 618\penalty0 (7965):\penalty0 500--505, 2023.

\bibitem[Kissinger and van~de Wetering(2019)]{pyzx}
Aleks Kissinger and John van~de Wetering.
\newblock Pyzx: Large scale automated diagrammatic reasoning.
\newblock \emph{arXiv preprint arXiv:1904.04735}, 2019.
\newblock URL \url{https://github.com/Quantomatic/pyzx/tree/master}.

\bibitem[Kitaev(1995)]{kitaev1995quantum}
A~Yu Kitaev.
\newblock Quantum measurements and the abelian stabilizer problem.
\newblock \emph{arXiv preprint quant-ph/9511026}, 1995.

\bibitem[Kremer et~al.(2024)Kremer, Villar, Paik, Duran, Faro, and Cruz-Benito]{kremer2024practical}
David Kremer, Victor Villar, Hanhee Paik, Ivan Duran, Ismael Faro, and Juan Cruz-Benito.
\newblock Practical and efficient quantum circuit synthesis and transpiling with reinforcement learning.
\newblock \emph{arXiv preprint arXiv:2405.13196}, 2024.

\bibitem[Lao and Browne(2022)]{lao20222qan}
Lingling Lao and Dan~E Browne.
\newblock {2QAN}: A quantum compiler for 2-local qubit hamiltonian simulation algorithms.
\newblock In \emph{Proceedings of the 49th Annual International Symposium on Computer Architecture}, pages 351--365, 2022.
\newblock URL \url{https://github.com/lllingoo/2QAN}.

\bibitem[Lattner and Adve(2004)]{lattner2004llvm}
Chris Lattner and Vikram Adve.
\newblock Llvm: A compilation framework for lifelong program analysis \& transformation.
\newblock In \emph{International symposium on code generation and optimization, 2004. CGO 2004.}, pages 75--86. IEEE, 2004.

\bibitem[Litteken et~al.(2023)Litteken, Seifert, Chadwick, Nottingham, Chong, and Baker]{litteken2023qompress}
Andrew Litteken, Lennart~Maximilian Seifert, Jason Chadwick, Natalia Nottingham, Frederic~T Chong, and Jonathan~M Baker.
\newblock Qompress: Efficient compilation for ququarts exploiting partial and mixed radix operations for communication reduction.
\newblock In \emph{Proceedings of the 28th ACM International Conference on Architectural Support for Programming Languages and Operating Systems, Volume 2}, pages 646--659, 2023.

\bibitem[Lloyd(1996)]{lloyd1996universal}
Seth Lloyd.
\newblock Universal quantum simulators.
\newblock \emph{Science}, 273\penalty0 (5278):\penalty0 1073--1078, 1996.

\bibitem[Majumdar et~al.(2023)Majumdar, Rivero, Metz, Hasan, and Wang]{majumdar:2023}
Ritajit Majumdar, Pedro Rivero, Friederike Metz, Areeq Hasan, and Derek~S. Wang.
\newblock {B}est practices for quantum error mitigation with digital zero-noise extrapolation.
\newblock \emph{arXiv preprint arXiv:2307.05203}, 2023.
\newblock \doi{10.48550/arXiv.2307.05203}.

\bibitem[Mariella and Zhuk(2023)]{dsm-swap}
Nicola Mariella and Sergiy Zhuk.
\newblock A doubly stochastic matrices-based approach to optimal qubit routing.
\newblock \emph{Quantum Information Processing}, 22\penalty0 (7):\penalty0 264, 2023.
\newblock URL \url{https://github.com/qiskit-community/dsm-swap}.

\bibitem[Matsakis and Klock(2014)]{10.1145/2663171.2663188}
Nicholas~D. Matsakis and Felix~S. Klock.
\newblock The {Rust} language.
\newblock In \emph{Proceedings of the 2014 ACM SIGAda Annual Conference on High Integrity Language Technology}, HILT '14, pages 103--104, New York, NY, USA, 2014. Association for Computing Machinery.
\newblock ISBN 9781450332170.
\newblock \doi{10.1145/2663171.2663188}.
\newblock URL \url{https://doi.org/10.1145/2663171.2663188}.

\bibitem[McClean et~al.(2020)McClean, Rubin, Sung, Kivlichan, Bonet-Monroig, Cao, Dai, Fried, Gidney, Gimby, et~al.]{mcclean2020openfermion}
Jarrod~R McClean, Nicholas~C Rubin, Kevin~J Sung, Ian~D Kivlichan, Xavier Bonet-Monroig, Yudong Cao, Chengyu Dai, E~Schuyler Fried, Craig Gidney, Brendan Gimby, et~al.
\newblock {OpenFermion}: the electronic structure package for quantum computers.
\newblock \emph{Quantum Science and Technology}, 5\penalty0 (3):\penalty0 034014, 2020.

\bibitem[McKinney et~al.(2023)McKinney, Hatridge, and Jones]{mckinney2023mirage}
Evan McKinney, Michael Hatridge, and Alex~K Jones.
\newblock {MIRAGE}: Quantum circuit decomposition and routing collaborative design using mirror gates.
\newblock \emph{arXiv preprint arXiv:2308.03874}, 2023.
\newblock URL \url{https://github.com/Pitt-JonesLab/mirror-gates}.

\bibitem[Miessen et~al.(2024)Miessen, Egger, Taverneilli, and Mazzola]{miessen:2024}
Alexander Miessen, Daniel~J Egger, Ivano Taverneilli, and Guglielmo Mazzola.
\newblock {B}enchmarking digital quantum simulations and optimization above hundreds of qubits using quantum critical dynamics.
\newblock \emph{arXiv preprint arXiv:2404.08053}, 2024.
\newblock \doi{10.48550/arXiv.2404.08053}.

\bibitem[Montanez-Barrera and Michielsen(2024)]{montanez:2024}
J.~A. Montanez-Barrera and Kristel Michielsen.
\newblock {T}owards a universal {QAOA} protocol: {E}vidence of quantum advantage in solving combinatorial optimization problems.
\newblock \emph{arXiv preprint arXiv:2405.09169}, 2024.
\newblock \doi{10.48550/arXiv.2405.09169}.

\bibitem[Mundada et~al.(2023)Mundada, Barbosa, Maity, Wang, Merkh, Stace, Nielson, Carvalho, Hush, Biercuk, et~al.]{mundada2023experimental}
Pranav~S Mundada, Aaron Barbosa, Smarak Maity, Yulun Wang, Thomas Merkh, TM~Stace, Felicity Nielson, Andre~RR Carvalho, Michael Hush, Michael~J Biercuk, et~al.
\newblock Experimental benchmarking of an automated deterministic error-suppression workflow for quantum algorithms.
\newblock \emph{Physical Review Applied}, 20\penalty0 (2):\penalty0 024034, 2023.

\bibitem[Murali et~al.(2019)Murali, Baker, Javadi-Abhari, Chong, and Martonosi]{murali2019noise}
Prakash Murali, Jonathan~M Baker, Ali Javadi-Abhari, Frederic~T Chong, and Margaret Martonosi.
\newblock Noise-adaptive compiler mappings for noisy intermediate-scale quantum computers.
\newblock In \emph{Proceedings of the twenty-fourth international conference on architectural support for programming languages and operating systems}, pages 1015--1029, 2019.

\bibitem[Murali et~al.(2020)Murali, McKay, Martonosi, and Javadi-Abhari]{murali2020software}
Prakash Murali, David~C McKay, Margaret Martonosi, and Ali Javadi-Abhari.
\newblock Software mitigation of crosstalk on noisy intermediate-scale quantum computers.
\newblock In \emph{Proceedings of the Twenty-Fifth International Conference on Architectural Support for Programming Languages and Operating Systems}, pages 1001--1016, 2020.

\bibitem[Nannicini et~al.(2022)Nannicini, Bishop, G{\"u}nl{\"u}k, and Jurcevic]{bip}
Giacomo Nannicini, Lev~S Bishop, Oktay G{\"u}nl{\"u}k, and Petar Jurcevic.
\newblock Optimal qubit assignment and routing via integer programming.
\newblock \emph{ACM Transactions on Quantum Computing}, 4\penalty0 (1):\penalty0 1--31, 2022.
\newblock URL \url{https://github.com/qiskit-community/qiskit-bip-mapper}.

\bibitem[Nielsen and Chuang(2001)]{nielsen2001quantum}
Michael~A Nielsen and Isaac~L Chuang.
\newblock Quantum computation and quantum information.
\newblock \emph{Phys. Today}, 54\penalty0 (2):\penalty0 60, 2001.

\bibitem[Patel et~al.(2008)Patel, Markov, and Hayes]{patel2008optimal}
Ketan~N Patel, Igor~L Markov, and John~P Hayes.
\newblock Optimal synthesis of linear reversible circuits.
\newblock \emph{Quantum Inf. Comput.}, 8\penalty0 (3):\penalty0 282--294, 2008.

\bibitem[Pelofske et~al.(2023)Pelofske, B\"{a}rtschi, Cincio, Golden, and Eidenbenz]{pelofske:2023}
Elijah Pelofske, Andreas B\"{a}rtschi, Lukasz Cincio, John Golden, and Stephan Eidenbenz.
\newblock {S}caling {W}hole-{C}hip {QAOA} for {H}igher-{O}rder {I}sing {S}pin {G}lass {M}odels on {H}eavy-{H}ex {G}raphs.
\newblock \emph{arXiv preprint arXiv:2312.00997}, 2023.
\newblock \doi{10.48550/arXiv.2312.00997}.

\bibitem[Peres and Galv{\~a}o(2023)]{peres2023quantum}
Filipa~CR Peres and Ernesto~F Galv{\~a}o.
\newblock Quantum circuit compilation and hybrid computation using {Pauli}-based computation.
\newblock \emph{Quantum}, 7:\penalty0 1126, 2023.

\bibitem[Peruzzo et~al.(2014)Peruzzo, McClean, Shadbolt, Yung, Zhou, Love, Aspuru-Guzik, and O'brien]{peruzzo2014variational}
Alberto Peruzzo, Jarrod McClean, Peter Shadbolt, Man-Hong Yung, Xiao-Qi Zhou, Peter~J Love, Al{\'a}n Aspuru-Guzik, and Jeremy~L O'brien.
\newblock A variational eigenvalue solver on a photonic quantum processor.
\newblock \emph{Nature communications}, 5\penalty0 (1):\penalty0 4213, 2014.

\bibitem[Peterson et~al.(2022)Peterson, Bishop, and Javadi-Abhari]{peterson2022optimal}
Eric~C Peterson, Lev~S Bishop, and Ali Javadi-Abhari.
\newblock Optimal synthesis into fixed {XX} interactions.
\newblock \emph{Quantum}, 6:\penalty0 696, 2022.

\bibitem[Puzzuoli et~al.(2023)Puzzuoli, Wood, Egger, Rosand, and Ueda]{qiskit-dynamics}
Daniel Puzzuoli, Christopher~J Wood, Daniel~J Egger, Benjamin Rosand, and Kento Ueda.
\newblock {Qiskit Dynamics}: A python package for simulating the time dynamics of quantum systems.
\newblock \emph{Journal of Open Source Software}, 8\penalty0 (90):\penalty0 5853, 2023.

\bibitem[Reinhold et~al.(2024)Reinhold, Teo, Jaskula, Chen, Thorgrimsson, Mishra, D'Ewart, Shaffer, Davis, Chen, Sivarajah, and Karalekas]{oqpy}
Philip Reinhold, Stephanie Teo, Jean-Christophe Jaskula, Li~Chen, Brandur Thorgrimsson, Anurag Mishra, Mitch D'Ewart, Ryan Shaffer, Erik Davis, Yi-Ting Chen, Prasahnt Sivarajah, and Peter Karalekas.
\newblock Oqpy: Openqasm 3 + openpulse in python, January 2024.
\newblock URL \url{https://doi.org/10.5281/zenodo.10534874}.

\bibitem[Robledo-Moreno et~al.(2024)Robledo-Moreno, Motta, Haas, Javadi-Abhari, Jurcevic, Kirby, Martiel, Sharma, Sharma, Shirakawa, Sitdikov, Sun, Sung, Takita, Tran, Yunoki, and Mezzacapo]{robledo:2024}
Javier Robledo-Moreno, Mario Motta, Holger Haas, Ali Javadi-Abhari, Petar Jurcevic, William Kirby, Simon Martiel, Kunal Sharma, Sandeep Sharma, Tomonori Shirakawa, Iskandar Sitdikov, Rong-Yang Sun, Kevin~J Sung, Maika Takita, Minh~C. Tran, Seiji Yunoki, and Antonio Mezzacapo.
\newblock {C}hemistry {B}eyond {E}xact {S}olutions on a {Q}uantum-{C}entric {S}upercomputer.
\newblock \emph{arXiv preprint arXiv:2405.05068}, 2024.
\newblock \doi{10.48550/arXiv.2405.05068}.

\bibitem[Seidel et~al.(2022)Seidel, Bock, Tcholtchev, and Hauswirth]{seidel2022qrisp}
Raphael Seidel, Sebastian Bock, Nikolay Tcholtchev, and Manfred Hauswirth.
\newblock Qrisp: A framework for compilable high-level programming of gate-based quantum computers.
\newblock \emph{PlanQC-Programming Languages for Quantum Computing}, 2022.

\bibitem[Seif et~al.(2024)Seif, Liao, Tripathi, Krsulich, Malekakhlagh, Amico, Jurcevic, and Javadi-Abhari]{seif2024suppressing}
Alireza Seif, Haoran Liao, Vinay Tripathi, Kevin Krsulich, Moein Malekakhlagh, Mirko Amico, Petar Jurcevic, and Ali Javadi-Abhari.
\newblock Suppressing correlated noise in quantum computers via context-aware compiling.
\newblock 2024.

\bibitem[Shinjo et~al.(2024)Shinjo, Seki, Shirakawa, Sun, and Yunoki]{shinjo:2024}
Kazuya Shinjo, Kazuhiro Seki, Tomonori Shirakawa, Rong-Yang Sun, and Seiji Yunoki.
\newblock {U}nveiling clean two-dimensional discrete time quasicrystals on a digital quantum computer.
\newblock \emph{arXiv preprint arXiv:2403.16718}, 2024.
\newblock \doi{10.48550/arXiv.2403.16718}.

\bibitem[Shtanko et~al.(2023)Shtanko, Wang, Zhang, Harle, Seif, Movassagh, and Minev]{shtanko:2023}
Oles Shtanko, Derek~S. Wang, Haimeng Zhang, Nikhil Harle, Alireza Seif, Ramis Movassagh, and Zlatko Minev.
\newblock Uncovering local integrability in quantum many-body dynamics.
\newblock \emph{arXiv preprint arXiv:2307.07552}, 2023.
\newblock \doi{10.48550/arXiv.2307.07552}.

\bibitem[Sivarajah et~al.(2020)Sivarajah, Dilkes, Cowtan, Simmons, Edgington, and Duncan]{tket}
Seyon Sivarajah, Silas Dilkes, Alexander Cowtan, Will Simmons, Alec Edgington, and Ross Duncan.
\newblock t$\ket{ket}$: a retargetable compiler for {NISQ} devices.
\newblock \emph{Quantum Science and Technology}, 6\penalty0 (1):\penalty0 014003, 2020.
\newblock URL \url{https://github.com/CQCL/tket}.

\bibitem[Smith et~al.(2021)Smith, Ravi, Murali, Baker, Earnest, Javadi-Abhari, and Chong]{smith2021error}
Kaitlin~N Smith, Gokul~Subramanian Ravi, Prakash Murali, Jonathan~M Baker, Nathan Earnest, Ali Javadi-Abhari, and Frederic~T Chong.
\newblock Error mitigation in quantum computers through instruction scheduling.
\newblock \emph{arXiv preprint arXiv:2105.01760}, 2021.

\bibitem[Smith(2017)]{smith2017someone}
Robert~S Smith.
\newblock Someone shouts, ``$\ket{01000}$!'' who is excited?
\newblock \emph{arXiv preprint arXiv:1711.02086}, 2017.

\bibitem[Smith et~al.(2020)Smith, Peterson, Skilbeck, and Davis]{quilc}
Robert~S Smith, Eric~C Peterson, Mark~G Skilbeck, and Erik~J Davis.
\newblock An open-source, industrial-strength optimizing compiler for quantum programs.
\newblock \emph{Quantum Science and Technology}, 5\penalty0 (4):\penalty0 044001, 2020.

\bibitem[Stavenger et~al.(2022)Stavenger, Crane, Smith, Kang, Girvin, and Wiebe]{stavenger2022c2qa}
Timothy~J Stavenger, Eleanor Crane, Kevin~C Smith, Christopher~T Kang, Steven~M Girvin, and Nathan Wiebe.
\newblock {C2QA}-bosonic qiskit.
\newblock In \emph{2022 IEEE High Performance Extreme Computing Conference (HPEC)}, pages 1--8. IEEE, 2022.

\bibitem[Steiger et~al.(2018)Steiger, H{\"a}ner, and Troyer]{projectq}
Damian~S Steiger, Thomas H{\"a}ner, and Matthias Troyer.
\newblock Project{Q}: an open source software framework for quantum computing.
\newblock \emph{Quantum}, 2:\penalty0 49, 2018.
\newblock URL \url{https://github.com/ProjectQ-Framework/ProjectQ}.

\bibitem[Suzuki et~al.(2021)Suzuki, Kawase, Masumura, Hiraga, Nakadai, Chen, Nakanishi, Mitarai, Imai, Tamiya, et~al.]{qiskit-qualcs}
Yasunari Suzuki, Yoshiaki Kawase, Yuya Masumura, Yuria Hiraga, Masahiro Nakadai, Jiabao Chen, Ken~M Nakanishi, Kosuke Mitarai, Ryosuke Imai, Shiro Tamiya, et~al.
\newblock Qulacs: a fast and versatile quantum circuit simulator for research purpose.
\newblock \emph{Quantum}, 5:\penalty0 559, 2021.
\newblock URL \url{https://github.com/Gopal-Dahale/qiskit-qulacs}.

\bibitem[Tang et~al.(2021)Tang, Tomesh, Suchara, Larson, and Martonosi]{tang2021cutqc}
Wei Tang, Teague Tomesh, Martin Suchara, Jeffrey Larson, and Margaret Martonosi.
\newblock {CutQC}: using small quantum computers for large quantum circuit evaluations.
\newblock In \emph{Proceedings of the 26th ACM International conference on architectural support for programming languages and operating systems}, pages 473--486, 2021.
\newblock URL \url{https://github.com/weiT1993/CutQC}.

\bibitem[Tao et~al.(2022)Tao, Shi, Yao, Li, Javadi-Abhari, Cross, Chong, and Gu]{tao2022giallar}
Runzhou Tao, Yunong Shi, Jianan Yao, Xupeng Li, Ali Javadi-Abhari, Andrew~W Cross, Frederic~T Chong, and Ronghui Gu.
\newblock Giallar: Push-button verification for the qiskit quantum compiler.
\newblock In \emph{Proceedings of the 43rd ACM SIGPLAN International Conference on Programming Language Design and Implementation}, pages 641--656, 2022.

\bibitem[Treinish et~al.(2022)Treinish, Carvalho, Tsilimigkounakis, and S{\'a}]{treinish:2021}
Matthew Treinish, Ivan Carvalho, Georgios Tsilimigkounakis, and Nahum S{\'a}.
\newblock {RustworkX}: A high-performance graph library for python.
\newblock \emph{Journal of Open Source Software}, 7\penalty0 (79):\penalty0 3968, 2022.
\newblock \doi{10.21105/joss.03968}.
\newblock URL \url{https://doi.org/10.21105/joss.03968}.

\bibitem[Van Den~Berg et~al.(2023)Van Den~Berg, Minev, Kandala, and Temme]{van2023probabilistic}
Ewout Van Den~Berg, Zlatko~K Minev, Abhinav Kandala, and Kristan Temme.
\newblock Probabilistic error cancellation with sparse {Pauli}--{Lindblad} models on noisy quantum processors.
\newblock \emph{Nature Physics}, pages 1--6, 2023.

\bibitem[Vazquez et~al.(2024)Vazquez, Tornow, Riste, Woerner, Takita, and Egger]{vazquez2024scaling}
Almudena~Carrera Vazquez, Caroline Tornow, Diego Riste, Stefan Woerner, Maika Takita, and Daniel~J Egger.
\newblock Scaling quantum computing with dynamic circuits.
\newblock \emph{arXiv preprint arXiv:2402.17833}, 2024.

\bibitem[Viola et~al.(1999)Viola, Knill, and Lloyd]{viola1999dynamical}
Lorenza Viola, Emanuel Knill, and Seth Lloyd.
\newblock Dynamical decoupling of open quantum systems.
\newblock \emph{Physical Review Letters}, 82\penalty0 (12):\penalty0 2417, 1999.

\bibitem[Wang et~al.(2022)Wang, Ding, Gu, Li, Lin, Pan, Chong, and Han]{hanruiwang2022quantumnas}
Hanrui Wang, Yongshan Ding, Jiaqi Gu, Zirui Li, Yujun Lin, David~Z Pan, Frederic~T Chong, and Song Han.
\newblock {QuantumNAS}: Noise-adaptive search for robust quantum circuits.
\newblock In \emph{The 28th IEEE International Symposium on High-Performance Computer Architecture (HPCA-28)}, 2022.
\newblock URL \url{https://github.com/mit-han-lab/torchquantum}.

\bibitem[Weaving et~al.(2023)Weaving, Ralli, Kirby, Tranter, Love, and Coveney]{weaving2023stabilizer}
Tim Weaving, Alexis Ralli, William~M Kirby, Andrew Tranter, Peter~J Love, and Peter~V Coveney.
\newblock A stabilizer framework for the contextual subspace variational quantum eigensolver and the noncontextual projection ansatz.
\newblock \emph{Journal of Chemical Theory and Computation}, 19\penalty0 (3):\penalty0 808--821, 2023.
\newblock URL \url{https://github.com/UCL-CCS/symmer/tree/main}.

\bibitem[Wei et~al.(2023)Wei, Lauer, Pritchett, Shanks, McKay, and Javadi-Abhari]{wei2023native}
Ken~Xuan Wei, Isaac Lauer, Emily Pritchett, William Shanks, David~C McKay, and Ali Javadi-Abhari.
\newblock Native two-qubit gates in fixed-coupling, fixed-frequency transmons beyond cross-resonance interaction.
\newblock \emph{arXiv preprint arXiv:2310.12146}, 2023.

\bibitem[Wittler et~al.(2021)Wittler, Roy, Pack, Werninghaus, Saha~Roy, Egger, Filipp, Wilhelm, and Machnes]{c3}
Nicolas Wittler, Federico Roy, Kevin Pack, Max Werninghaus, Anurag Saha~Roy, Daniel~J. Egger, Stefan Filipp, Frank~K. Wilhelm, and Shai Machnes.
\newblock Integrated tool set for control, calibration, and characterization of quantum devices applied to superconducting qubits.
\newblock \emph{Physical Review Applied}, 15\penalty0 (3), Mar 2021.
\newblock \doi{10.1103/physrevapplied.15.034080}.
\newblock URL \url{https://github.com/q-optimize/c3}.

\bibitem[Yasuda et~al.(2023)Yasuda, Suzuki, Kubota, Nakajima, Gao, Zhang, Shimono, Nurdin, and Yamamoto]{yasuda:2023}
Toshiki Yasuda, Yudai Suzuki, Tomoyuki Kubota, Kohei Nakajima, Qi~Gao, Wenlong Zhang, Satoshi Shimono, I.~Nurdin, Hendra, and Naoki Yamamoto.
\newblock {Q}uantum reservoir computing with repeated measurements on superconducting devices.
\newblock \emph{arXiv preprint arXiv:2310.06706}, 2023.
\newblock \doi{10.48550/arXiv.2310.06706}.

\bibitem[Yu et~al.(2023)Yu, Zhao, Wei, et~al.]{yu2023simulating}
Hongye Yu, Yusheng Zhao, Tzu-Chieh Wei, et~al.
\newblock Simulating large-size quantum spin chains on cloud-based superconducting quantum computers.
\newblock \emph{Physical Review Research}, 5\penalty0 (1):\penalty0 013183, 2023.

\bibitem[Zhang et~al.(2021)Zhang, Hayes, Qiu, Jin, Chen, and Zhang]{toqm}
Chi Zhang, Ari~B Hayes, Longfei Qiu, Yuwei Jin, Yanhao Chen, and Eddy~Z Zhang.
\newblock Time-optimal qubit mapping.
\newblock In \emph{Proceedings of the 26th ACM International Conference on Architectural Support for Programming Languages and Operating Systems}, pages 360--374, 2021.
\newblock URL \url{https://github.com/qiskit-community/qiskit-toqm}.

\bibitem[Zhang et~al.(2023)Zhang, Wu, Wang, Li, Shapourian, Shabani, and Ding]{zhang2023oneq}
Hezi Zhang, Anbang Wu, Yuke Wang, Gushu Li, Hassan Shapourian, Alireza Shabani, and Yufei Ding.
\newblock {OneQ}: A compilation framework for photonic one-way quantum computation.
\newblock In \emph{Proceedings of the 50th Annual International Symposium on Computer Architecture}, pages 1--14, 2023.

\bibitem[Zhang and Nation(2023)]{zhang:2023}
Victoria Zhang and Paul~D. Nation.
\newblock {C}haracterizing quantum processors using discrete time crystals.
\newblock \emph{arXiv preprint arXiv:2301.07625}, 2023.
\newblock \doi{10.48550/arXiv.2301.07625}.

\end{thebibliography}

\clearpage
\appendix
\onecolumngrid
\section{Suppressing errors in dynamic circuits}\label{sec:error-suppression}

In this appendix we discuss some experimental details about the dynamic circuit experiment presented in the main text. Specifically, we discuss how real-time classical control flow, which is a characteristic of dynamic circuits, are executed on the quantum computer. We then use that knowledge to suppress some known sources of coherent errors.

\begin{figure*}[h]
  \includegraphics[width=\textwidth]{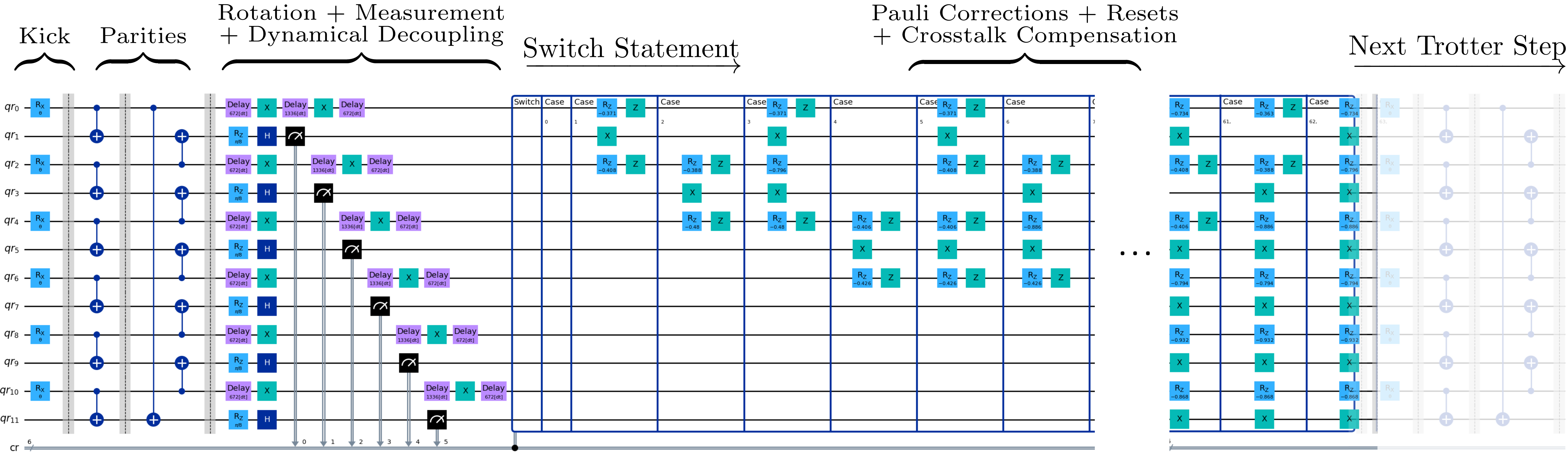}
  \caption{Suppressing errors in the simulation of a honeycomb Ising lattice using dynamic circuits. Long latencies can be problematic as they contribute to long periods of decoherence and crosstalk. We suppress them by performing dynamical decoupling in parallel with measurements, and designing a {\tt switch-case} statement that not only performs the circuit requirements for Pauli correction and ancilla reset, but also compensates for the accumulated crosstalk depending on the state of the ancilla. One Trotter step is depicted, whereas error suppression allows us to extend to 8 Trotter steps in the experiment.}
  \label{fig:dynamicexp}
\end{figure*}

The details of our dynamic circuit experiment are shown in Figure~\ref{fig:dynamicexp}. The circuit works by allocating every other qubit as ancilla. The data are entangled with ancilla in two CNOT layers. This computes the parity of each pair of data (the original sites of the honeycomb Ising lattice) into an ancilla located in-between that pair. The next layer consists of rotating and measuring out the ancilla. Since measurements are long ($1400 \, \text{ns}$ on this device), it is critical that we perform dynamical decoupling in this layer to suppress the effects of decoherence on the data qubits.

The next layer of the circuit is the Pauli correction performed on data qubits, conditioned on the outcome of previous ancilla measurements. Again, given that feed-forward latencies can be long ($1060 \, \text{ns}$ on this device), it is important that we suppress errors as much as possible. However, on the generation of control electronics available in this experiment, no quantum gates can be carried out while the feed-forward instructions are being evaluated, rendering dynamical decoupling challenging. This creates a long period of time where not only qubit decoherence, but also correlated errors in the form of crosstalk are present. The latter are unitary errors, which can be quadratically more detrimental if not dealt with effectively. Fortunately, their unitary nature also means that we can counter them in the rest of the circuit.

To suppress correlated errors, we first transform the circuit from using a series of {\tt if} statements to using a single {\tt switch-case} statement. This is again due to the characteristics of the specific control system where each {\tt if} statement is evaluated and acted upon in series, leading to a long accumulation of decoherence and static crosstalk on the qubits. In contrast, a {\tt switch} statement is evaluated once, and one and only one {\tt case} is executed after that evaluation. This leads to a circuit with constant duration regardless of which branch is selected at runtime.

We now have to build all the {\tt case}s such that maximal suppression is achieved. We know the round-trip time between when classical measurement outcomes (bits) are made available to the controller, to the time the {\tt switch} statement is evaluated, to when the next instruction is executed. Due to static crosstalk between neighboring qubits, single-qubit $Z$ and two-qubit $ZZ$ rotations can accumulate on the qubits (this is due to the convention of crosstalk as the Hamiltonian $H_{\ket{11}\bra{11}} = \frac{\nu}{2} (Z\otimes Z - I\otimes Z - Z\otimes I$)). 

In our circuit we can use the fact that every other qubit is an ancilla, and is measured. If the qubit is measured in the ground state, then crosstalk is already suppressed. Alternately if the ancilla is measured in the excited state, then two-qubit $ZZ$ is transformed to a single-qubit $Z$. All in all, this means that we can compensate for the crosstalk during the {\tt switch} statement by inserting a conditional inverse $Z$ rotation. We provision for this compensation within each branch, according to its {\tt case}~\cite{seif2024suppressing}. 

Lastly, all ancilla must be reset before the next Trotter step. Again, rather than using a $reset$ instruction which can itself be long, we rely on the fact that within each {\tt case} we know the state of all ancillas. Therefore reset is yet another conditional gate in the branch, this time a conditional $X$ gate. These suppression techniques combine to yield the good performance of dynamic circuits in the main text.

\section{ISA for IBM Quantum Heron Processors}
\label{sec:isa}

A concrete example of an Instruction Set Architecture (ISA) for a quantum processor is described here. The latest generation of IBM Quantum processor as of this writing is the Heron family of devices with 133-qubits and a tunable coupler architecture that allows for faster, lower-error two-qubits gates while simultaneously reducing crosstalk from static $ZZ$ interactions. The instructions supported on Heron processors are listed in Table~\ref{tab:heron_isa} along with syntax examples of each in OpenQASM format. Note that RZ and DELAY require classical arguments. The RZ instruction accepts an angle argument, $\theta$. DELAY accepts a time argument, $\tau$. The data types for these classical arguments depend on the serialization format. In OpenQASM, $\theta$ is of type \texttt{angle}, which is a special-purpose fixed-point representation of angles. The DELAY argument $\tau$, when represented in OpenQASM, should be a \texttt{duration} type with units of ``\texttt{dt}'' and must be an integer.

\begin{table}[h]
    \centering
    \begin{tabular}{lccl}
    \hline
        Instruction & Operands & Syntax example & Semantics \\
        \hline
        RZ($\theta$) & \texttt{angle} $\theta$, $q_0$ & \texttt{rz($\theta$) q[0];} & Continuous rotation about Pauli-Z, i.e. $\exp(-iZ\theta/2)$\\
        SX & $q$ & \texttt{sx q[0];} & $\pi/2$ rotation about Pauli-X, i.e. $\exp(-iX\pi/4) = \sqrt{\mathrm{X}}$\\
        X & $q_0$ & \texttt{x q[0];} & Pauli-X gate\\
        DELAY($\tau$) & \texttt{duration} $\tau$, $q_0$ & \texttt{delay(4dt) q[0];} & Injects an idle operation with a specific duration\\
        MEASURE & $q_0$ & \texttt{c = measure q[0];} & Measure $q_0$ and return a classical bit\\
        RESET & $q_0$ & \texttt{reset q[0];} & Reset $q_0$ to state $\ket{0}$. Used mid-circuit.\\
        INIT & $q_0$ & \texttt{init q[0];} & Initialize $q_0$ in state $\ket{0}$.\\
        CZ & $q_0$, $q_1$ & \texttt{cz q[0], q[1];} & Controlled-Z gate\\
        BARRIER & $q_0$, [$q_1$, ...] & \texttt{barrier q[0], q[1];} & Prevents movement of instructions across the barrier\\
        IF/ELSE & boolean expression & \texttt{if (c[0] || c[1] == 1)} & Branching directives\\
        SWITCH & controlling value & \texttt{switch(c) \{case 0 \{\} \}} & Branches controlled by a common value\\
        FOR & iteration expression & \texttt{for i in [0:9]} & For loop\\
        WHILE & boolean expression & \texttt{while (c[0] != 0)} & While loop\\
        \hline
    \end{tabular}
    \caption{Instructions of the Heron ISA. Operands column refers to \texttt{qubit} arguments with the shorthand $q_i$. OpenQASM data types are provided for classical arguments. Syntax examples are in OpenQASM format.}
    \label{tab:heron_isa}
\end{table}

Some instructions are restricted in the allowed set of operands. For instance, the CZ gate requires two quantum operands $q_0$ and $q_1$, and the edge $(q_0, q_1)$ must appear in the connectivity graph of the Heron processor.

RESET and INIT both reset qubits to the $\ket{0}$ state, but INIT is only permitted as the first instruction in a quantum circuit, and is implemented with a slower mechanism that produces higher-fidelity initial states. RESET may be used anywhere in a circuit and has a more balanced trade-off between speed and fidelity. The Heron processor will maximally parallelize operations on independent qubits. The BARRIER instruction creates bounded regions to constrain the start and end times of instructions between barriers.

Heron processors also admit use of control flow instructions. Conditions in IF, ELSE, FOR, and WHILE statements may include comparisons with operators: $==$, $!=$, $>=$, $<=$, $>$, $<$, as well as classical expressions on bits involving the bit-wise arithmetic operations: AND (\texttt{\&}), OR (\texttt{|}), and XOR (\texttt{\^}).

The current Heron ISA does not specify timing data associated with classical arithmetic expressions or from setup and branching in control flow instructions. Thus, there is an inherent uncertainty in exact timing of operations in circuits containing these operations. The Heron ISA also does not constrain what operations may be nested inside control-flow blocks, which introduces sufficient runtime complexity that the current execution model is forced to serialize all control flow. This significantly hampers the ability to deliver short-duration dynamic circuits, which are critical to extracting the potential performance improvements made possible by introducing measurement and feed-forward in circuits.

Future versions of the Heron ISA will add constraints to the use of control flow to permit a more powerful execution model with parallel execution of independent control-flow blocks. In particular, branches from IF, ELSE, and SWITCH statements will be scheduled such that all branches have equal duration so that knowledge of which-path information does not need to be shared in the control system to retain tight synchronization. Nested control-flow statement involving runtime values will not be permitted, because it complicates lifetime analysis of values. In this constrained model, FOR and WHILE must either not have control flow inside or they must be countable loops, meaning loops with trip counts that are known at compile-time.

Even with these added constraints, users still desire the ability to modify circuits to suppress errors using dynamical decoupling or the kind of manipulation described in Sec.~\ref{sec:error-suppression}. However, it is difficult to deliver precise timing information related to classical expressions or control flow because the particular timing may depend on contextual details, such as whether or not a classical register is available in the controller at time of execution of an instruction. Thus, rather than extending the Heron ISA with additional timing data, we intend to add \texttt{stretch}es as introduced in the OpenQASM spec as a mechanism to allow users to describe timing intent without access to exact timing data.

These additional constraints on dynamic circuits will be critical to enhancing circuit performance on Heron and other near-term IBM quantum processors. We anticipate being able to relax most, or even all, of these constraints in the fault-tolerant era because the slower effective clock speed of logical operations is compatible with a more complex execution model.

\end{document}